\documentclass[a4paper]{article}

\usepackage{graphicx}
\usepackage{amssymb,amsfonts,amsmath}
\usepackage[round,colon]{natbib}

\newcounter{firstbib}

\begin{document}


{\Large\bf Robust estimation of microbial diversity in theory and in practice}

\bigskip

Bart Haegeman$^1$, J\'er\^ome Hamelin$^2$, John Moriarty$^3$, Peter Neal$^4$, Jonathan Dushoff$^5$, Joshua S.\ Weitz$^6$

\smallskip

$^1$\textit{Centre for Biodiversity Theory and Modelling, Experimental Ecology Station, Centre National de Recherche Scientifique, Moulis, France;}
$^2$\textit{INRA, UR50, Laboratoire de Biotechnologie de l'Environnement, Narbonne, France;}
$^3$\textit{School of Mathematics, University of Manchester, Manchester, United Kingdom;}
$^4$\textit{Department of Mathematics and Statistics, University of Lancaster, Lancaster, United Kingdom;}
$^5$\textit{Department of Biology and Institute of Infectious Disease Research, McMaster University, Hamilton, Ontario, Canada;}
$^6$\textit{School of Biology and School of Physics, Georgia Institute of Technology, Atlanta, Georgia, United States of America}

\smallskip

\begin{quote}
\textbf{Quantifying diversity is of central importance for the study of structure, function and evolution of microbial communities. The estimation of microbial diversity has received renewed attention with the advent of large-scale metagenomic studies. Here, we consider what the diversity observed in a sample tells us about the diversity of the community being sampled. First, we argue that one cannot reliably estimate the absolute and relative number of microbial species present in a community without making unsupported assumptions about species abundance distributions. The reason for this is that sample data do not contain information about the number of rare species in the tail of species abundance distributions. We illustrate the difficulty in comparing species richness estimates by applying Chao's estimator of species richness to a set of \emph{in silico} communities: they are ranked incorrectly in the presence of large numbers of rare species. Next, we extend our analysis to a general family of diversity metrics (``Hill diversities''), and construct lower and upper estimates of diversity values consistent with the sample data. The theory generalizes Chao's estimator, which we retrieve as the lower estimate of species richness. We show that Shannon and Simpson diversity can be robustly estimated for the \emph{in silico} communities. We analyze nine metagenomic data sets from a wide range of environments, and show that our findings are relevant for empirically-sampled communities. Hence, we recommend the use of Shannon and Simpson diversity rather than species richness in efforts to quantify and compare microbial diversity.}

Accepted paper in press at \textit{The ISME Journal};  doi:10.1038/ismej.2013.10

\textbf{Subject category:} Microbial population and community ecology

\end{quote}

\newpage

\section*{Introduction}

Species diversity is a crucial property of ecological communities: it is the primary descriptor of community structure, and it is generally believed to be a major determinant of the functioning and the dynamics of ecological communities \citep{Wilson1999, Loreau2001, Ives2007, Loreau2010}.  Therefore, diversity measurement is often a first step in characterizing an ecological community \citep{Brose2003, Magurran2004, Gotelli2011}.  Because an exhaustive census of the community is usually not feasible, community diversity must be inferred from the diversity observed in a sample taken from the community.  The inference problem can be difficult, especially when community diversity is believed to be very large~\citep{Engen1978,Bunge1993,Mao2005}.

Diversity measurement is particularly challenging for microbial communities \citep{Hughes2001, Bohannan2003, Kemp2004, Schloss2005, Sloan2008, Bunge2009, Ovreas2011}.  First, it should be recalled that there is no unambiguous way to define microbial ``species'' \citep{Stackebrandt2002}.  Here we use the term species pragmatically to mean an operationally determined taxonomic unit (e.g., 97\% identity of 16S rRNA \citep{Schloss2005}).  However measured, the species diversity of microbial communities is usually much larger than that of communities of larger organisms.  Moreover, the number of organisms in microbial communities is typically many orders of magnitude larger than the number of organisms in plant or animal communities \citep{Whitman1998}.  This leads to severe sampling problems.  Although metagenomic approaches allow for impressively large sample size \citep{Huber2007, Roesch2007, Rusch2007}, even these huge samples correspond to a tiny fraction of the community being sampled.  Hence, for microbial community samples, community diversity is generally much larger than sample diversity.  This disparity between community and sample leads to a challenge that we address here: how can microbial diversity be estimated robustly?

One popular approach to circumvent the sampling problem is to assume that the species abundance distribution of the community belongs to a specific family (for example, the family of lognormal distributions) \citep{Curtis2002, Hong2006, Schloss2006, Quince2008}.  Such an assumption fills in the information about the community missing in the data and leads to precise diversity estimates.  But the validity of the estimates depends crucially on the choice of the species abundance distribution family.  This choice cannot be verified empirically because the sample data do not contain sufficient information about the community structure.  In fact, many distribution families yield extrapolated community structures that are consistent with the sample data.  Here we show that the extrapolation approach has intrinsic limitations.

Other methods for diversity estimation have been proposed.  For example, proposals have been made to extrapolate the rarefaction curve beyond the actual sample size \citep{Gotelli2001, Colwell2004}, or to assume a particular distribution for the community diversity over taxonomic levels \citep{May1988, Mora2011}.  Eventually, also these methods are limited by the lack of information about the community structure in the sample data.  Rather than filling this gap by unverifiable assumptions, here we ask what can (and cannot) be inferred from the sample data alone.  An interesting step in this direction is given by the popular Chao estimator \citep{Chao1984, Shen2003, Chao2009}.  Chao's estimate can be interpreted as a \emph{lower estimate} of the species richness consistent with the data.  We take the estimation strategy underlying Chao's estimator a step further, and construct lower and upper estimates for a general family of community diversities, including species richness, Shannon diversity and Simpson diversity \citep{Hill1973}.  The unification we propose here represents a robust approach to estimating microbial diversity in theory and in practice.

\section*{Materials and methods}

\subsubsection*{Data sets}  The data sets used in this paper were downloaded from the supplementary material of \citet{Quince2008}.  The abundance data used in Figure~1 correspond to 16S rDNA sequences obtained from a bacterial soil community (sample ``Brazil'' in \citet{Roesch2007}).  The abundance data used in Figure~5 correspond to 16S rDNA sequences obtained from a bacterial seawater community from the upper ocean \citep{Rusch2007}, from four bacterial soil communities \citep{Roesch2007}, and from bacterial and archaeal seawater communities from two hydrothermal vents \citep{Huber2007}.

\subsubsection*{Rank-abundance curves}  We represent the species abundance distribution of a community as a rank-abundance curve, that is, we arrange the species in decreasing order of community abundance, and plot species abundance as a function of species rank.  We use logarithmic scales for both axes of the rank-abundance curves, so that a community with power-law abundance distribution is represented as a straight line (the slope is equal to the power-law exponent), see Figure~2A.  We constructed the communities of Figure~1 by using a piecewise linear parametrization of the rank-abundance curve.  Hence, the species abundance distributions consist of power-law segments with different exponents.

\subsubsection*{Rarefaction curves}  We define $S_m$ as the expected number of species in a sample of $m$ individuals taken from the community (sampling with replacement).  The rarefaction curve of the community is the plot of the number of species $S_m$ as a function of the sample size $m$.  It is important to distinguish the community rarefaction curve from the rarefaction curve estimated from sample data.  For a sample of size $M$ taken from the community, the part of the rarefaction curve corresponding to $S_m$ with $m\leq M$ can be estimated by subsampling the sample data.  The same approach fails for the part of the rarefaction curve corresponding to $S_m$ with $m>M$.  In that case the rarefaction curve has to be extrapolated, introducing large estimation uncertainty.  We studied two extreme extrapolation scenarios:  one for the slowest (i.e., smallest slope) and one for the fastest (i.e., largest slope) increase of the rarefaction curve compatible with the sample data, see Figure~3.

\subsubsection*{Hill diversities}  The Hill diversities, defined in Equation~(\ref{eq:hilldef}), can be computed if the community abundances are known.  If only sample data are available, Hill diversities have to be estimated.  We consider sampling with replacement, and denote by $M$ the sample size and by $F_k$ the number of species sampled $k$ times.  We developed an estimation procedure that exploits the link between Hill diversities $D_\alpha$ and the rarefaction curve $S_m$.  The lower estimate $\widehat S_m^-$ of the rarefaction curve,
\begin{equation*}
 \widehat S_m^- = \begin{cases}
 \sum_{k \geq 1} F_k \Big( 1 - \frac{\binom{M-k}{m}}{\binom{M}{m}} \Big)
 & \text{if $m \leq M$} \\
 S_\text{obs} + \frac{F_1^2}{2 F_2}
 \Big( 1 - \big( 1 - \frac{2 F_2}{M F_1} \big)^{m-M} \Big)
 & \text{if $m > M,$}
 \end{cases}
\end{equation*}
yields the lower estimate of the Hill diversity,
\begin{equation}
 \widehat D_\alpha^- = \bigg( \sum_{m=1}^\infty
 \frac{\alpha\;\Gamma(m-\alpha)}{m!\;\Gamma(1-\alpha)}\,
 \widehat S_m^- \bigg)%
 ^{\frac{1}{1-\alpha}},
 \label{eq:dalphamin}
\end{equation}
where $\Gamma$ denotes the gamma function.  Similarly, the upper estimate of the rarefaction curve,
\begin{equation*}
 \widehat S_m^+ = \begin{cases}
 \sum_{k \geq 1} F_k \Big( 1 - \frac{\binom{M-k}{m}}{\binom{M}{m}} \Big)
 & \text{if $m \leq M$} \\
 S_\text{obs} + \frac{N F_1}{M}
 \Big( 1 - \big( 1 - \frac{1}{N} \big)^{m-M} \Big)
 & \text{if $m > M,$}
 \end{cases}
\end{equation*}
with $N$ the (estimated) community size, yields the upper estimate of the Hill diversity,
\begin{equation}
 \widehat D_\alpha^+ = \bigg( \sum_{m=1}^\infty
 \frac{\alpha\;\Gamma(m-\alpha)}{m!\;\Gamma(1-\alpha)}\,
 \widehat S_m^+ \bigg)%
 ^{\frac{1}{1-\alpha}}.
 \label{eq:dalphaplus}
\end{equation}
The estimators (\ref{eq:dalphamin}) and (\ref{eq:dalphaplus}) can be computed with the Matlab code in the Supplementary Information, and were used to generate Figures~4 and 5.

\section*{Results}

\subsubsection*{Species richness cannot be estimated from sample data alone}

\begin{figure}
\begin{center}
\includegraphics[width=.95\textwidth]{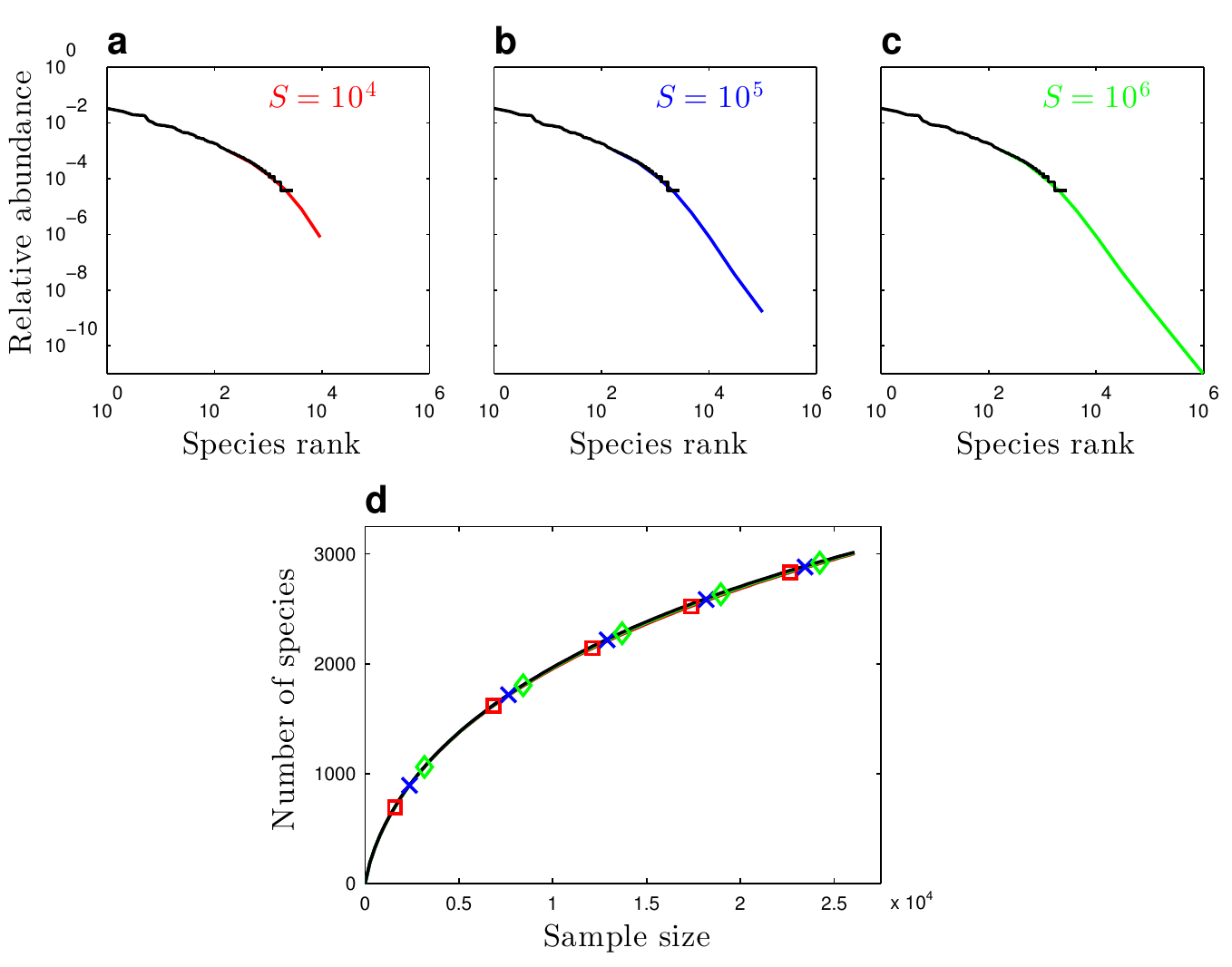}
\end{center}
\caption{Empirical sample data are consistent with very different communities.  We consider the abundance data of a sample taken from a bacterial soil community  (sample ``Brazil'' in \citet{Roesch2007}).  The sample consists of $26079$ individuals belonging to $2880$ species.  We tried to reconstruct the community from which the sample was taken.  Panels~a--c show the rank-abundance curve of three such reconstructed communities.  The first community (panel~a, in red) has $10^4$ species;  the second community (panel~b, in blue) has $10^5$ species;  the third community (panel~c, in green) has $10^6$ species.  For each of the three reconstructions the community rank-abundance curve is an extension of the sample rank-abundance curve (in black).  We claim that each of the three reconstructed communities is compatible with the sample data.  This can be seen from the rarefaction curves in panel~d:  the rarefaction curve for the sample data (black line) coincides with the rarefaction curves for the reconstructed communities (red line with squares for community in panel a, blue line with $\times$-marks for community in panel b, and green line with diamonds for community in panel c).  Because the sample data are consistent with very different values of the community richness, the community richness cannot be estimated from the sample data.} \label{fig:fig1}
\end{figure}

We are interested in estimating the diversity of a community based on the composition of a sample taken from the community.  Our approach is to reconstruct community structures, i.e., species abundance distributions, from the sample data.  For the example data set of Figure~1, we find that a wide range of communities are consistent with the sample data.  The reconstructed communities have vastly different numbers of species, differing by two orders of magnitude, implying that estimating species richness is subject to large biases.

We claim that sample data is always consistent with very different community structures.  To establish this claim we study the link between the rare species tail of the community and the sample data, summarized by the rarefaction curve.  A computation in Supplementary Text~S1 shows that the rarefaction curve up to sample size $M$ is insensitive to the abundance distribution of species with relative abundance well below $\frac{1}{M}$.  For concreteness we set a relative abundance threshold at $\frac{1}{50\,M}$, and we call the species with larger and smaller relative abundance than this threshold the ``non-rare'' and ``rare'' species, respectively.  The computation shows that the rarefaction curves does not depend on the abundance distribution of the rare species.  Changes in the rare species tail, such as increasing the number of rare species by several orders of magnitude (but keeping the total abundance of rare species constant), does not affect the sample data.  As a consequence, estimating species richness is intrinsically problematic.

Note that we use a statistical definition of rarity which depends on the sampling effort $M$; the set of rare species gets smaller when sampling gets deeper.  This contrasts with the ecological concept of rarity, a community property independent of sample size \citep{PedrosAlio2006, Sogin2006}, see the Discussion section.

To further illustrate the theoretical result we reconsider the reconstructed communities of Figure~1.  The communities have the same abundance distribution of the non-rare species.  In each community the set of rare species occupies $0.5 \%$ of the total community abundance, explaining why the corresponding rarefaction curves coincide, see Figure~1D.  Nevertheless, the number of rare species differs by two orders of magnitude.  Another example of \emph{in silico} communities with very different rare species tails but with the same rarefaction curve is shown in Supplementary Figure~S1.

We conclude that sample data do not allow us to distinguish communities with very different rare species tails.  The insensitivity of the rarefaction curve to rare species implies that it is difficult or impossible to reliably estimate the community species richness from sample data alone.

\subsubsection*{Relative species richness cannot be estimated from sample data alone}

\begin{figure}
\begin{center}
\includegraphics[width=.9\textwidth]{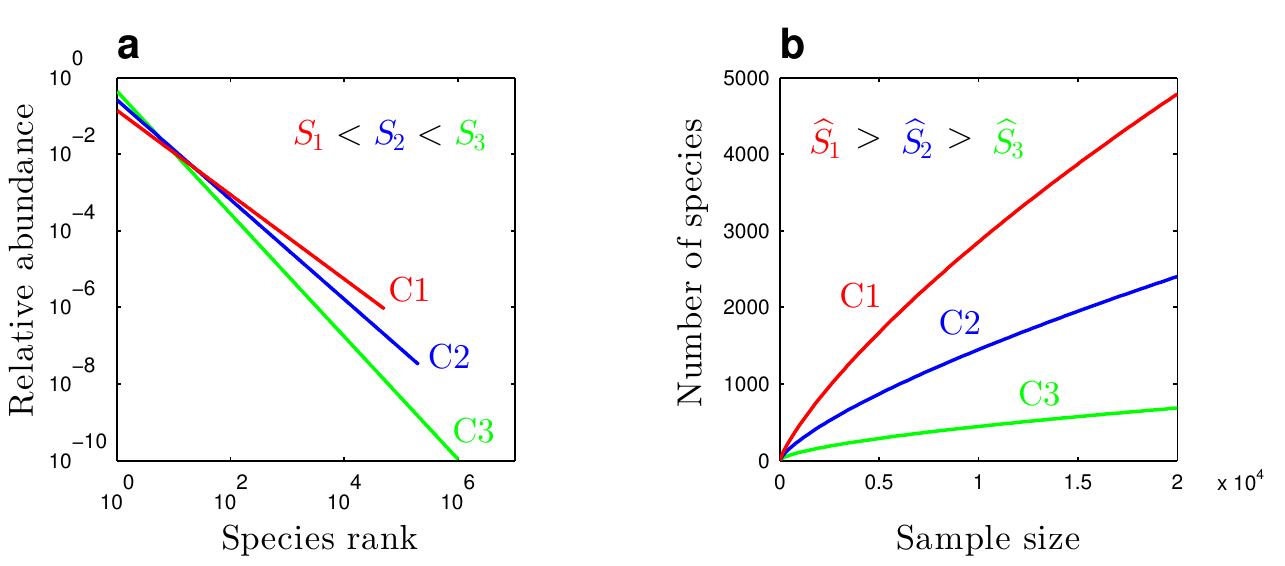}
\end{center}
\caption{Estimated species richness does not rank correctly communities.  We generated three community abundance distributions, the rank-abundance curves of which are shown in panel~a.  Community C1 (red) has the smallest number of species; community C3 (green) has the largest number of species.  The rarefaction curves of the three communities up to sample size $2\;10^4$ are shown in panel~b.  Based on the rarefaction data, one would conclude that community C1 is the most diverse and community C3 the least diverse.  Hence, the ranking of the communities according to their observed diversity is inverted compared to the ranking according to their true diversity.  This observation is confirmed when applying Chao's estimator to sample data.  Community C1 is estimated to have 10 times more species than community C3, whereas in reality community C1 has 20 times less species than community C3.  See Supplementary Table~S1 for the numerical data of the communities.} \label{fig:fig2}
\end{figure}

We have shown that the number of species in a community cannot be reliably estimated from sample data.  A related question is whether sample data can be used to rank different communities according to their number of species.  In this section we show that this cannot be done without additional assumptions.

We present an explicit example to illustrate the use of sample data to rank communities, see Figure~2.  We consider three communities which differ widely in species richness:  community C1 has 20 times fewer species than community C3.  We construct the initial arcs of these rarefaction curves, see Figure~2B.  Surprisingly, the rarefaction curves suggest that community C1 is the most diverse, and community C3 the least diverse.  We therefore expect that any estimator of species richness ranks the communities in the inverse order of their true species richness.  Indeed, Chao's estimator predicts that community C1 has almost 10 times as many species as community C3 (see Supplementary Table~S1; values are averaged over sample randomness).

To understand the incorrect ranking we take a closer look at the communities in Figure~2A.  We explained, in the previous section, that sample data are insensitive to rare species.  When we compare the number of non-rare species in the communities (species with relative abundance above $10^{-6}$), we find that community C1 has 15 times more non-rare species than community C3.  This explains why the sample data suggest that community C1 is the most diverse.  Community C1 has a large number of non-rare species combined with a relatively small number of rare species.  In contrast, community C3 has a relatively small number of non-rare species combined with a very large number of rare species.  This explains the discrepancy between true number of species, mainly determined by the rare species, and estimated number of species, determined by the non-rare species.

The example of Figure~2 indicates a general problem:  relative species richness cannot be reliably estimated.  The problem is due to the same mechanism as the one identified in the previous section.  Sample data cannot be used to rank communities according to their number of species because sample data do not contain information about the number of rare species.

\subsubsection*{Some generalized diversities can be estimated from sample data alone}

Altough insensitive to rare species, sample data do contain information about the community structure.  In this section we demonstrate that diversity indices that are weakly dependent on rare species can be estimated from sample data.

Diversity is a broader notion than species richness.  Alternative definitions of diversity have been proposed in which rare species contribute less than common species.  These alternative diversities account not only for species richness but also for the evenness of the community structure.  Examples are the Shannon diversity index \citep{Shannon1948} and the Simpson diversity index \citep{Simpson1949}.  Here we study a family of generalized diversities, the Hill diversities $D_\alpha$ \citep{Hill1973} that includes these two examples as well as species richness as special cases.  For a community consisting of $S$ species with relative abundances $p_1,p_2,\ldots,p_S$, the Hill diversities are defined by
\begin{equation}
D_\alpha = \left( \sum_{i=1}^S p_i^\alpha \right)^{\frac{1}{1-\alpha}}.
\label{eq:hilldef}
\end{equation}
We obtain a Hill diversity for each value of the parameter $\alpha$.  For $\alpha=0$ the species are weighted equally in the sum of Equation~(\ref{eq:hilldef}) (each term is equal to one), and $D_0=S$, i.e., $D_0$ is equal to species richness.  For $\alpha>0$ the species are not weighted equally.  Instead, a rare species contributes less than a common species.  For larger values of $\alpha$ the weighting is more unequal, see Supplementary Text~S2. As an extreme case, only the most abundant species contributes in the limit $\alpha\to\infty$.  The Hill diversity of order 1 is related to the Shannon diversity index (note that Definition~(\ref{eq:hilldef}) should be understood as $D_1 = \lim_{\alpha\to 1} D_\alpha$) and the Hill diversity of order 2 is related to the Simpson concentration index.  The Hill diversity for a community in which all $S$ species have the same relative abundance $p_i = \frac{1}{S}$ is equal to $D_\alpha = S$ for any value of the parameter $\alpha$.  This indicates that any Hill diversity $D_\alpha$ can be considered as an effective number of species \citep{Hill1973,Jost2006}, which facilitates the interpretation of estimated diversity values and allows us to compare the estimation properties of different Hill diversities.

As $\alpha$ increases the Hill diversities are increasingly insensitive to the tail of rare species and are more strongly determined by the non-rare species, see Supplementary Figure~S2.  Hence, we expect that they are more accurately estimated from sample data.  A mathematical link between the Hill diversities and the rarefaction curve further indicates which Hill diversities can be estimated from sample data.  In Supplementary Text~S3 we show that any Hill diversity $D_\alpha$ can be expressed in terms of the rarefaction curve.  The Hill diversity $D_2$ is related to the initial slope of the rarefaction curve \citep{Lande2000}.  Thus, for $\alpha$ close to 2, the Hill diversity $D_\alpha$ depends on the part of the rarefaction curve for small sample size.  For smaller $\alpha$, the Hill diversity $D_\alpha$ depends on the rarefaction curve for increasingly large sample size.  The Hill diversity $D_0$ is equal to species richness, which can be obtained as the limit of the rarefaction curve for infinite sample size.

These observations have important implications for the diversity estimation problem.  We suppose that sample data of size $M$ are given, and we try to estimate the rarefaction curve at sample size $m$.  The community rarefaction curve for sample sizes $m\leq M$ can be estimated in an unbiased manner by subsampling the sample data, but for $m > M$ the rarefaction curve can only be estimated based on extrapolation. This leads to increasingly biased estimates as $m$ increases.  Hence, we reach the following conclusions.  On one hand, Hill diversities that depend on the initial part of the rarefaction curve, that is, $D_\alpha$ for $\alpha$ close to 2, can be estimated robustly.  On the other hand, Hill diversities that depend on the part of the rarefaction curve for large sample size, that is, $D_\alpha$ for $\alpha$ close to 0, cannot be estimated robustly.  We now seek to make this classification of community diversities more precise.

\subsubsection*{Estimators for Hill diversities}

\begin{figure}
\begin{center}
\includegraphics[width=.54\textwidth]{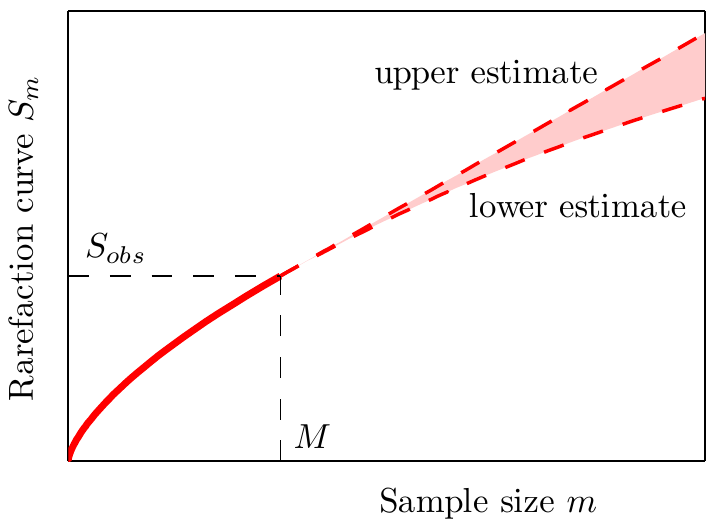}
\end{center}
\caption{Extrapolating the rarefaction curve.  The Hill diversity estimators $\widehat D_\alpha^-$ and $\widehat D_\alpha^+$ are based on reconstructions of the rarefaction curve $S_m$ from sample data.  For a sample of size $M$, the rarefaction curve $S_m$ for $m \leq M$ can be estimated by subsampling (red full line).  If the sample size $M$ is large, the estimator has small uncertainty.  The rarefaction curve $S_m$ for $m > M$ can be estimated by extrapolating the sample data beyond the sample size $M$.  Different extrapolation scenarios are compatible with the sample data.  We consider two extreme scenarios (red dashed lines).  A lower estimate is obtained by assuming that unobserved species are approximately as rare as the rarest observed species.  An upper estimate is obtained by assuming that unobserved species are represented in the community by one individual.  The difference between the two extremes quantifies the uncertainty of the extrapolation, shown as the red shaded region.  The uncertainty increases rapidly for $m \gg M$.}
\label{fig:fig3}
\end{figure}

We have argued that the Hill diversities $D_\alpha$ with $\alpha$ close to 2 can be estimated accurately, and that the Hill diversities $D_\alpha$ with $\alpha$ close to 0 cannot be estimated accurately.  In this section we introduce and study estimators for the set of Hill diversities $D_\alpha$ with $0\leq\alpha\leq 2$.

We have shown that a wide variety of communities may be consistent with any given sample data.  Here we look for two extreme members of this set of reconstructed communities.  We construct a lower estimate of the diversity, $\widehat D_\alpha^-$, by assuming that unobserved species are approximately as rare as the rarest observed species.  We construct an upper estimate of the diversity, $\widehat D_\alpha^+$, by assuming that unobserved species are represented in the community by a single individual.  We first extrapolate the rarefaction curve based on these assumptions, see Figure~3, and then use the extrapolated curves to calculate the Hill diversities.  The detailed construction of the estimators $\widehat D_\alpha^-$ and $\widehat D_\alpha^+$ is presented in Supplementary Texts~S3, S4 and S5.  A summary of the estimator formulas can be found in the Materials and Methods section.  We provide Matlab code to compute the estimators in the Supplementary Information.

Two properties follow directly from the definition of the estimators $\widehat D_\alpha^-$ and $\widehat D_\alpha^+$, see Supplementary Text~S5.  First, the lower estimate $\widehat D_0^-$ for species richness is equal to Chao's estimator.  Hence, the lower estimate $\widehat D_\alpha^-$ generalizes Chao's estimator for Hill diversities $D_\alpha$ with $\alpha > 0$.  Second, the estimators for Simpson diversity $D_2$ coincide, $\widehat D_2^- = \widehat D_2^+$.  This corresponds to the existence of an unbiased, non-parametric estimator for the Simpson concentration index, and confirms that Simpson diversity $D_2$ is particularly easy to estimate, even for small sample size $M$.  Note that the lower estimate can be computed from the sample data alone, but the upper estimate also requires an estimate of the community size $N$.

\begin{figure}
\begin{center}
\includegraphics[width=.8\textwidth]{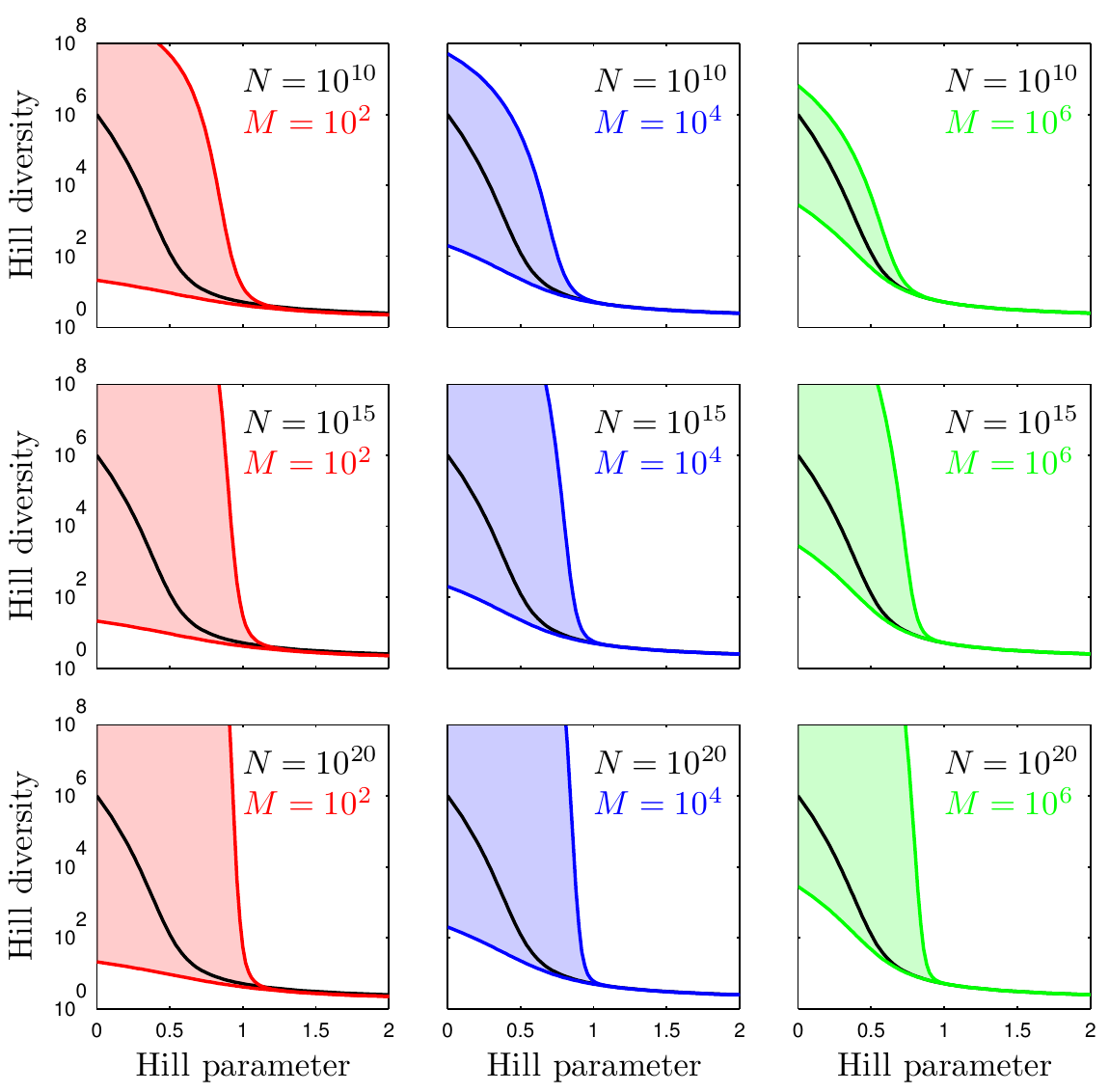}
\end{center}
\caption{Estimated Hill diversities for \emph{in silico} communities.  We generated samples from a community with power-law abundance distribution ($S=10^6$, $z=2$) and evaluated the estimators $\widehat D_\alpha^+$ and $\widehat D_\alpha^-$ for the Hill diversity $D_\alpha$.  We consider three sample sizes $M$ (in columns: $M=10^2, 10^4, 10^6$) and three community sizes $N$ (in rows: $N=10^{10}, 10^{15}, 10^{20}$).  The shaded range between $\widehat D_\alpha^+$ and $\widehat D_\alpha^-$ indicates the estimation uncertainty.  The true Hill diversity $D_\alpha$ of the community is plotted in black.  The Hill diversities between $\alpha=1$ (Shannon) and $\alpha=2$ (Simpson) are correctly estimated even for small sample size $M$.  The estimates of Hill diversities less than $\alpha = 1$, including $\alpha = 0$ (species richness), are characterized by large uncertainty.}
\label{fig:fig4}
\end{figure}

In Figure~4 we apply the estimators $\widehat D_\alpha^-$ and $\widehat D_\alpha^+$ to sample data from an \emph{in silico} community.  For $\alpha>1$ the lower and upper estimates almost coincide, so that the Hill diversities $D_\alpha$ with $\alpha>1$, and in particular Simpson diversity $D_2$, may be estimated with small error.  This holds for any sample size $M$ (as small as $M=100$) and any community size $N$.  For $\alpha<1$ the upper estimate increases steeply, so that the estimation uncertainty of the Hill diversities $D_\alpha$ with $\alpha$ small, and in particular species richness $D_0$, is very large.  This holds for any sample size $M$ (as large as $M=10^6$) and any community size $N$ much greater than $M$.  The effect of sample size $M$ and community size $N$ is only pertinent for $\alpha$ close to 1.  For these values of $\alpha$ the range between the lower and upper estimates narrows with increasing sample size $M$ and decreasing community size $N$, so that increasingly accurate estimates are obtained for Shannon diversity $D_1$.

\begin{figure}
\begin{center}
\includegraphics[width=.8\textwidth]{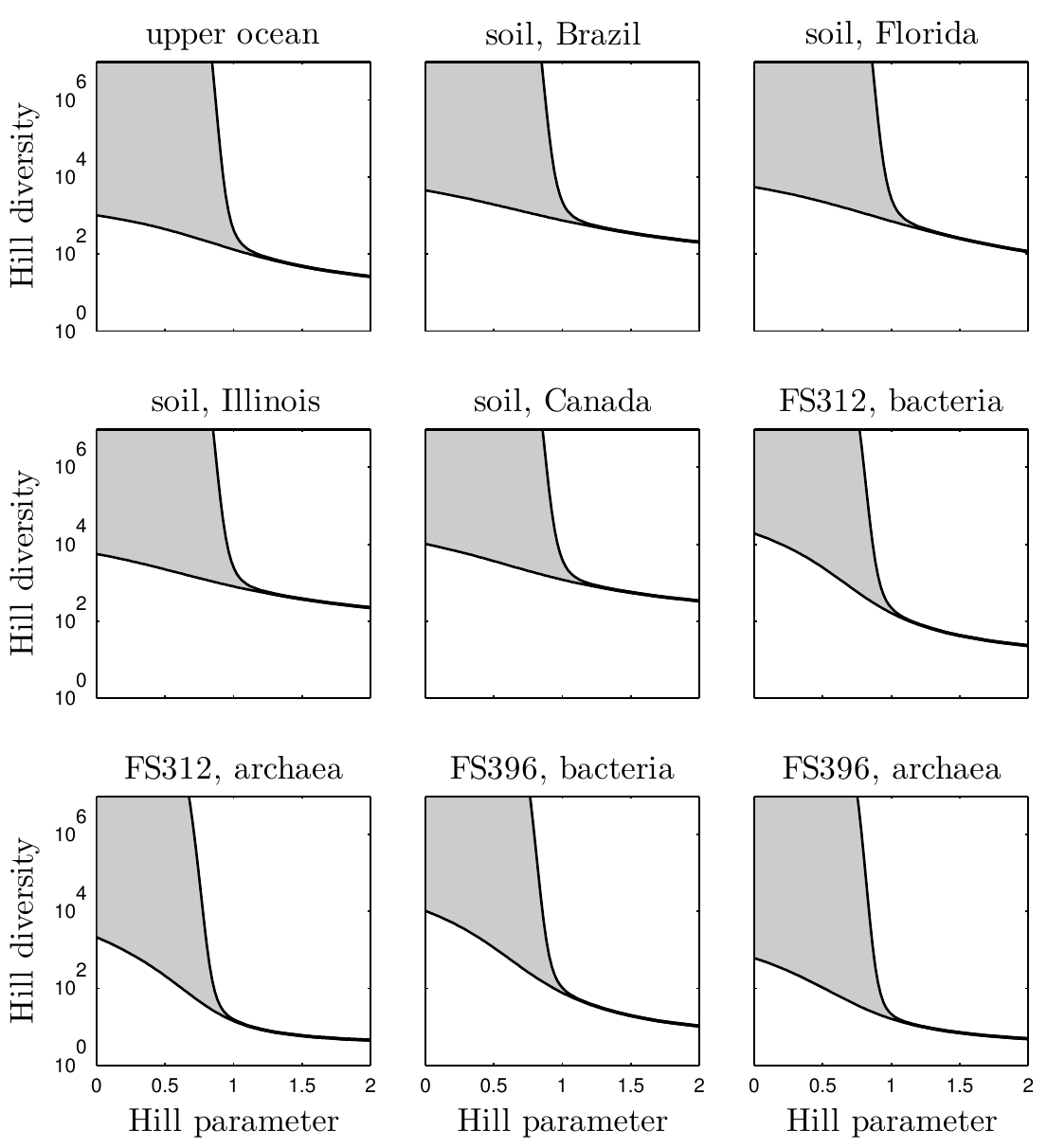}
\end{center}
\caption{Estimated Hill diversities for natural microbial communities.  We observe the same behavior as for the \emph{in silico} generated data sets of Figure~4: for $\alpha\geq 1$ the Hill diversity $D_\alpha$ can be estimated accurately; for $\alpha<1$ the estimation of the Hill diversity $D_\alpha$ has large uncertainty.  We used the same data sets as \citet{Quince2008}: a seawater bacterial sample from the upper ocean \citep{Rusch2007}, soil bacterial samples at four locations: Brazil, Florida, Illinois and Canada \citep{Roesch2007}, and seawater samples from deep-sea vents at two locations: FS312 and FS396, separated into bacteria and archaea \citep{Huber2007}.  The community size was set to $N=10^{15}$ for illustration;  results are robust to changes in community size (see Supplementary Figure~S4).}
\label{fig:fig5}
\end{figure}

We observe the same behavior when applying the Hill diversity estimators to empirical sample data, see Figure~5.  We applied the estimators to nine metagenomic data sets from a wide range of environments:  soil samples at four locations \citep{Roesch2007}, a seawater sample from the upper ocean \citep{Rusch2007} and seawater samples at two deep-sea vent locations \citep{Huber2007}.  The estimators exhibit the same patterns as for the \emph{in silico} community studied in Figure~4.  The Hill diversities $D_\alpha$ for $\alpha \geq 1$, including Shannon and Simpson diversity, can be estimated reliably.  For small $\alpha$ the estimation uncertainty is very large, that is, Hill diversities close to species richness cannot be estimated reliably.  The dependence of the estimation accuracy on the (estimated) community size $N$ is weak, see Supplementary Figure~S4.  These observations show that our analysis for \emph{in silico} communities is relevant for real communities as well.

\section*{Discussion}

We have argued that the estimation of species richness is intrinsically problematic.  We have provided evidence in three different but related ways.  First, we have shown that it is possible to add a large number of rare species to the community without significantly affecting its statistical properties under fixed-size sampling, see Figure~1.  As the number of added rare species can be large, the estimation uncertainty of the number of species is large as well.  Second, we have discussed an exact relationship between the community rarefaction curve and the set of Hill diversities.  Hill diversities close to Simpson's are based on the initial part of the rarefaction curve, which can be reliably interpolated from sample data.  Hill diversities beyond Shannon's, and species richness in particular, depend on parts of the rarefaction curve orders of magnitude beyond the actual sample size, whose estimation requires unverifiable extrapolation.  Third, we have constructed two estimators related to the Hill diversities, delimiting the range in which each true Hill diversity is expected to lie.  This range is relatively narrow for diversities from Simpson's to Shannon's, but it diverges for diversities towards species richness, see Figures~4 and 5.  Hence, the estimation uncertainty of species richness is intrinsically large.

We have also studied a weaker form of species richness estimation, namely, whether communities can be ranked according to species richness based on sample data.  We have argued that also in this case the sample data are not sufficiently informative.  The example shown in Figure~2 is interesting, because the community ranking based on estimated species richness, although completely different from the ranking based on true richness, is the same as the ranking based on true Simpson or Shannon diversity, see Supplementary Table~S1.  This observation can be understood intuitively. The insensitivity of the species richness estimator to the very rare species in the community is shared by the Simpson and Shannon diversity, but not by the community species richness.  In fact, different diversity estimators often yield the same community ranking \citep{Shaw2008}.  This should not be interpreted as an indication of the validity of the ranking for species richness;  the ranking based on true species richness can be completely different.  Communities should only be ranked according to community properties that can be estimated reliably.

The intrinsic problem of species richness estimation can be unlocked by introducing more information in the estimation procedure.  Obviously, the reliability of the estimate crucially depends on the reliability of the additional information.  For example, assuming a family of abundance distributions (for example, lognormal) can lead to species richness estimates with small uncertainty \citep{Schloss2005, Hong2006, Quince2008}.  But both the estimate and the uncertainty are conditional on the assumed distribution family.  In particular, assuming a species abundance distribution also fixes the rare species tail and, as we have argued, the sample data contain little information about the rare species tail.  Hence, the choice of distribution family is arbitrary.  Still, this choice strongly affects the species richness estimate.  We believe this to be a serious problem for this approach to diversity estimation.

Other assumptions have been introduced to make diversity estimation manageable.  Some regularity has been observed in the distribution of diversity over coarse taxonomic groups \citep{Mora2011}.  This regularity can be assumed down to the species level to guide the estimation of species richness.  Clearly, the approach depends crucially on the unverifiable validity of the extrapolation.  More generally, this and other approaches attempt to reduce the wide range of diversity values consistent with the data to a single value.  This implies that the reduction step is based on detailed information not contained in the sample data.  Such an approach is necessarily very sensitive to the detailed assumptions, and therefore not robust.

\citet{Mao2005} pointed out that rare species pose a serious problem for estimating species richness.  In this paper we have shown a practical way forward by quantifying the range of diversity values consistent with the data.  The latter idea underlies our construction of lower and upper estimates of community diversity, and is also crucial for Chao's estimator of species richness \citep{Chao1984}.  This estimator does not attempt to directly assess true species richness, but rather approximates the lowest species richness consistent with the sample data.  In many practical cases this indirect estimation is the most informative claim that can be made about species richness.

Different studies have highlighted the role of rare species in microbial communities \citep{Dykhuizen1998, PedrosAlio2006, Sogin2006, PedrosAlio2007, Huber2007, Gobet2010}.  We have argued that sample data contain limited information about the rare species tail of the community.  For example, the total number of rare species cannot be estimated.  However, an estimator for the relative abundance of unobserved species is available, see Supplementary Text~S4.  For the data sets we have analyzed the estimated relative abundance ranges from 0.1\% to 5\%, see Supplementary Table~S2.  These estimates depend on sample size.  It might be more practical to use a notion of rarity that is independent of sample size.  For example, we could call a species rare if its community abundance is below a certain threshold value (for example, relative abundance below $10^{-4}$).  We plan to address the problem of estimating the relative abundance of rare species in a sample-independent fashion as part of future work.

In this paper we have only considered taxonomic diversity.  Other notions of diversity such as functional and phylogenetic diversity are becoming increasingly popular \citep{HornerDevine2006, Lozupone2007, Green2008}.  Our study suggests that any diversity metrics that strongly depend on rare species will be difficult or impossible to estimate robustly.  It is interesting to note that other measurement techniques for microbial diversity are confronted with limitations similar to those of the sample-based techniques discussed in this paper.  The reassociation kinetics of community DNA are affected by community diversity \citep{Torsvik1990, Gans2005}, but it has been argued that not species richness, but Simpson and Shannon diversity can be estimated from the data \citep{Haegeman2008}.  Fingerprinting techniques provide snapshots of the community structure \citep{Fromin2002}: in this context also, the estimation of species richness seems to be impossible for highly diverse communities \citep{Loisel2006, Bent2008}, but preliminary results indicate that accurate estimators can be constructed for Simpson diversity.  Estimates of the total number of genes in a species, i.e., the pan genome size, has been estimated from a small number of sample genomes \citep{Tettelin2005}, but it is has been argued that these estimates are not robust and that similarity-based metrics should be used instead \citep{Kislyuk2011}.

These findings together with those of this paper make a strong case for the versatility of generalized diversities for the analysis of microbial diversity estimation.  They can be interpreted as effective number of species giving greater weight to common species \citep{Hill1973,Jost2006}, and have superior estimation properties compared to species richness.  We recommend the use of Shannon and Simpson diversity to quantify and compare microbial taxonomic diversity.

\section*{Acknowledgments}

Financial support for B.H. and J.H was provided by the DISCO project from the French National Research Agency (ANR, project number AAP215-SYSCOMM-2009), and for B.H, J.H., J.M. and P.N. by an Alliance grant from the British Council and the French Foreign Affairs Ministry (project number 22732SJ).  J.S.W. holds a Career Award at the Scientific Interface from the Burroughs Wellcome Fund.


\newpage

\renewcommand{\thesection}{S\arabic{section}}
\setcounter{section}{0}
\renewcommand{\theequation}{S\arabic{equation}}
\setcounter{equation}{0}
\renewcommand{\thefigure}{S\arabic{figure}}
\setcounter{figure}{0}
\renewcommand{\thetable}{S\arabic{table}}
\setcounter{table}{0}

\begin{center}
{\LARGE Supplementary Information} \\
\bigskip \bigskip
{\LARGE\bf Robust estimation of microbial diversity \\[3pt] in theory and in practice} \\
\bigskip \bigskip \bigskip
{\large B.~Haegeman, J.~Hamelin, J.~Moriarty, P.~Neal, J.~Dushoff, J.~S.~Weitz}
\end{center}

\bigskip \bigskip

\noindent {\bf Supplementary Text}
\begin{enumerate}
\item[] Text S1 \ Contribution of rare species to rarefaction curve
\item[] Text S2 \ Contribution of rare species to Hill diversities
\item[] Text S3 \ Hill diversities and rarefaction curve
\item[] Text S4 \ Estimating species abundances from sample data
\item[] Text S5 \ Estimating Hill diversities from sample data
\end{enumerate}

\bigskip

\noindent {\bf Supplementary Tables}
\begin{enumerate}
\item[] Table S1 \ Description of communities used in Figure~2
\item[] Table S2 \ Data for empirically-sampled microbial communities
\end{enumerate}

\bigskip

\noindent {\bf Supplementary Figures}
\begin{enumerate}
\item[] Figure S1 \ Sample data are insensitive to rare species tail of community
\item[] Figure S2 \ Hill diversity for large $\alpha$ is insensitive to rare species tail
\item[] Figure S3 \ Rank-abundances curve of empirical microbial community samples
\item[] Figure S4 \ Community-size dependence of Hill diversity estimates
\end{enumerate}

\bigskip

\noindent {\bf Computer code}
\begin{enumerate}
\item[] Matlab code to compute Hill diversity estimates
\end{enumerate}

\newpage

\noindent {\bf\LARGE Supplementary Text}

\section*{Text S1} 
\subsection*{Contribution of rare species to rarefaction curve}
\bigskip

We define $S_m$ as the expected number of species in a sample of $m$ individuals taken from the community.  The rarefaction curve of the community is the plot of the number of species $S_m$ as a function of the sample size $m$.  We consider a community consisting of $S$ species with relative abundance $p_1,p_2,\ldots,p_S$.  Then the expected number of sampled species $S_m$ is given by
\begin{equation}
S_m = \sum_{i=1}^S \Big( 1-(1-p_i)^m \Big).
\label{eq:rarefact}
\end{equation}

It is important to distinguish the community rarefaction curve (\ref{eq:rarefact}) from the rarefaction curve estimated from sample data.  We consider a sample of size $M$ taken from the community.  We denote the number of species observed in the sample by $S_\text{obs}$, and the number of species with abundance $k$ in the sample by $F_k$.  For $m\leq M$ the rarefaction curve $S_m$ can be estimated by taking subsamples of size $m$ out of the sample.  The average number of species observed in the subsample (averaged over all subsamples of size $m$) is an estimator for $S_m$,
\begin{equation}
\widehat S_m = \sum_{k \geq 1} F_k \bigg( 1 -
\frac{\binom{M-k}{m}}{\binom{M}{m}} \bigg),
\qquad m\leq M.
\label{eq:rfactestim}
\end{equation}
This estimator is reliable in the sense that it is unbiased (that is, the expected value of $\widehat S_m$ is equal to $S_m$).  Moreover, there is no other unbiased estimator with smaller variance.  For $m > M$ the estimation of the rarefaction curve is necessarily based on extrapolation, leading to less reliable estimates, especially for $m\gg M$.

We define a species to be rare if its relative abundance is much smaller than $\frac{1}{M}$.  This means that a rare species is unlikely to be present in the sample (of size $M$).  For concreteness we say that
\begin{equation}
\text{species $i$ is rare}
\quad \text{if} \quad
p_i \leq \frac{1}{50\,M}.
\label{eq:defrare}
\end{equation}
Note that our definition of rarity depends on the sample size $M$.  The choice of a threshold for rarity is arbitrary, though our results are robust to changes in the constant (which in this case has been set to 50) so long as it is much greater than 1.

We consider the rarefaction curve (\ref{eq:rarefact}) up to sample size $M$.  The contribution of species $i$ can be written as
\begin{equation*}
1-(1-p_i)^m = \sum_{j=1}^m \binom{m}{j} p_i^j\,(1-p_i)^{m-j},
\qquad m\leq M.
\end{equation*}
The $j$-th term in this sum is the probability that species $i$ is represented $j$ times in a sample of size $m$.  For a rare species $i$ we have $p_i \ll \frac{1}{M} \leq \frac{1}{m}$, and the first term dominates the other terms.  Hence,
\begin{equation*}
1-(1-p_i)^m \approx m\,p_i\,(1-p_i)^{m-1} \approx m\,p_i,
\qquad m\leq M.
\end{equation*}
Partitioning the set of species into rare and non-rare species, we get
\begin{align}
S_m &\approx \sum_{\substack{i=1 \\ i\ \text{non-rare}}}^S
           \hspace{-2ex} \Big( 1-(1-p_i)^m \Big) \hspace{1ex}
         + \sum_{\substack{i=1 \\ i\ \text{rare}}}^S m\,p_i \notag \\
        &= \sum_{\substack{i=1 \\ i\ \text{non-rare}}}^S
           \hspace{-2ex} \Big( 1-(1-p_i)^m \Big) \hspace{1ex}
         + m\,p_\text{rare},
\qquad m\leq M,
\label{eq:rfapprox}
\end{align}
with $p_\text{rare}$ the total relative abundance of the set of rare species in the community.

\begin{figure}[!ht]
\begin{center}
\includegraphics[width=.85\textwidth]{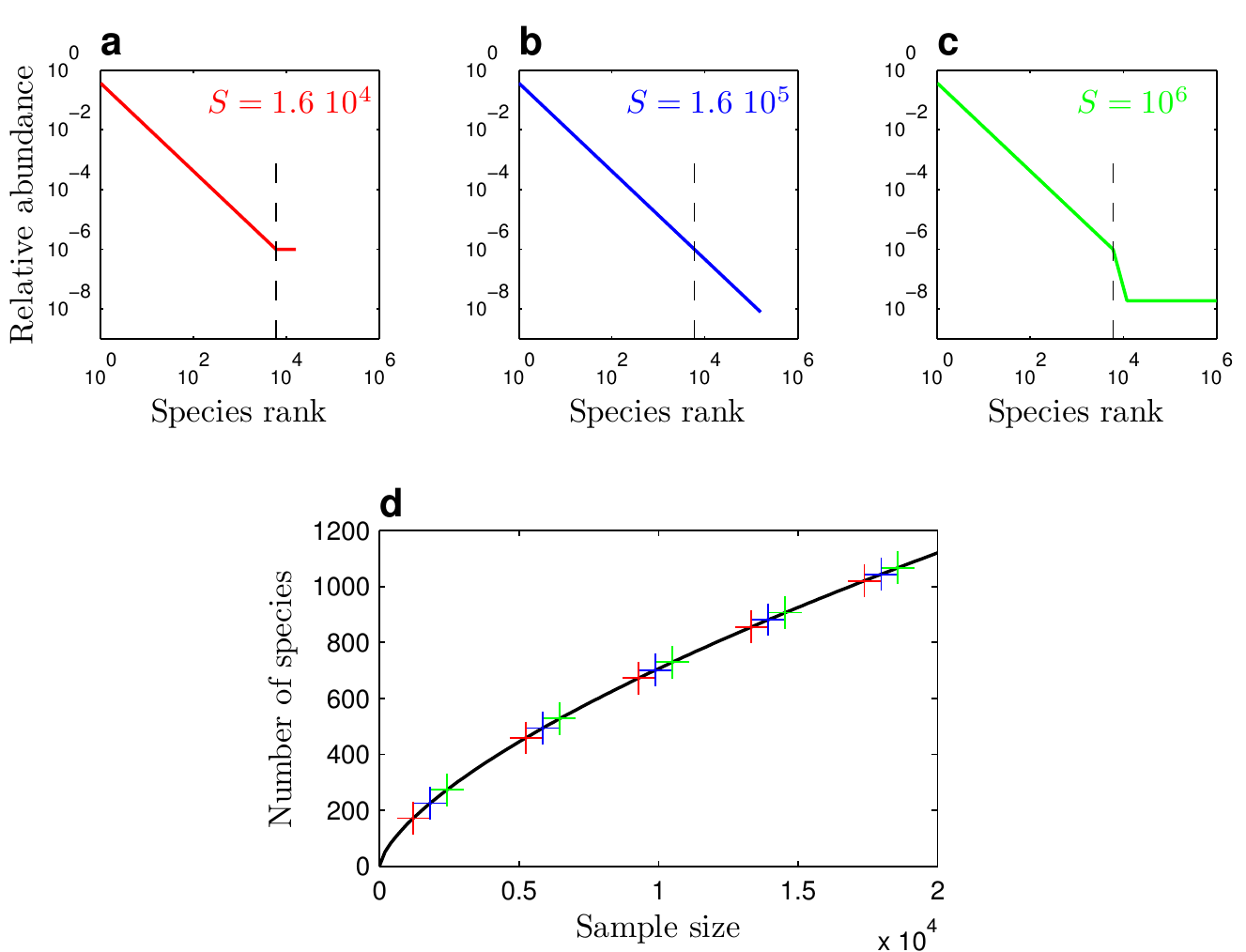}
\end{center}
\caption{Sample data are insensitive to rare species tail of community.  We generated three community abundance distributions, shown in red, blue and green (panels~a--c).  The three communities have the same abundance distribution for species with relative abundance above $10^{-6}$ (the part of the rank-abundance curve to the left of the dashed black line).  This common part consists of $6\;10^3$ species, occupying $99\%$ of the community abundance.  The communities differ in the tail of rare species:  the community in panel~a has $1.6\;10^4$ species;  the community in panel~b has $1.6\;10^5$ species;  the community in panel~c has $10^6$ species.  Despite the marked differences, the rarefaction curves of the three communities up to sample size $2\;10^4$ are identical (see panel~d).  This observation holds generally:  any set of rare species leads to the same rarefaction curve if each rare species has relative abundance below $10^{-6}$ and the total relative abundance of the set of rare species equals $0.01$.\label{fig:figS1}}
\end{figure}

From Equation~(\ref{eq:rfapprox}) it follows that the rarefaction curve does not depend on the abundance distribution of the rare species, but only on the total abundance of the rare species.  This follows directly from Definition (\ref{eq:defrare}): it is unlikely that a rare species will be observed twice in a sample of size $m$ (when $m<M$).  Therefore, the contribution of the rare species to the sample species richness depends only on their prevalence in the sample which, in turn, depends only on their prevalence in the community.  In particular, rarefaction curves obtained for different abundance distributions of the rare species are indistinguishable, see Figure~\ref{fig:figS1}.

\newpage
\section*{Text S2} 
\subsection*{Contribution of rare species to Hill diversities}
\bigskip

In the main text we have introduced the Hill diversities $D_\alpha$,
\begin{equation}
D_\alpha = \left( \sum_{i=1}^S p_i^\alpha \right)^{\frac{1}{1-\alpha}}.
\label{eq:defhill}
\end{equation}
The Hill diversity of order 1 is defined as the limit $D_1 = \lim_{\alpha\to 1} D_\alpha$, and is related to the Shannon diversity index $H$,
\begin{equation}
D_1 = \mathrm{e}^H
\qquad \text{with} \qquad
H = \sum_{i=1}^S -p_i \ln p_i.
\label{eq:shannon}
\end{equation}
The Hill diversity of order 2 is related to the Simpson concentration index $C$,
\begin{equation*}
D_2 = \frac{1}{C}
\qquad \text{with} \qquad
C = \sum_{i=1}^S p_i^2.
\end{equation*}
The Hill diversity of order $\infty$ is related to the relative abundance $p_\text{max}$ of the most abundant species,
\[
D_\infty = \frac{1}{p_\text{max}}
\qquad \text{with} \qquad
p_\text{max} = \max \big\{p_1,p_2,\ldots,p_S\big\}.
\]

We consider a community in which the rare species occupy a fraction $p_\text{rare}$ of the total community abundance.  We study the dependence of the Hill diversity on the number of rare species $S_\text{rare}$.  Assuming that the rare species have equal abundance, we get
\begin{align}
D_\alpha &= \Bigg(
\sum_{\substack{i=1\\i\ \text{non-rare}}}^S
\hspace{-2ex} p_i^\alpha \,+
\sum_{\substack{i=1\\i\ \text{rare}}}^S
p_i^\alpha \hspace{2ex} \Bigg)^{\frac{1}{1-\alpha}} \nonumber \\
&= \Bigg(
\sum_{\substack{i=1\\i\ \text{non-rare}}}^S
\hspace{-2ex} p_i^\alpha \,+\, S_\text{rare}\,
\Big( \frac{p_\text{rare}}{S_\text{rare}} \Big)^\alpha
\;\Bigg)^{\frac{1}{1-\alpha}} \nonumber \\
&= \Bigg(
\sum_{\substack{i=1\\i\ \text{non-rare}}}^S
\hspace{-2ex} p_i^\alpha \,+\, p_\text{rare}^\alpha\,
S_\text{rare}^{1-\alpha}\ \Bigg)^{\frac{1}{1-\alpha}}.
\label{eq:hillrare}
\end{align}
The first term inside the brackets contains the contribution of the non-rare species.  The second term inside the brackets, $p_\text{rare}^\alpha\, S_\text{rare}^{1-\alpha}$, contains the contribution of the rare species.  The contribution of the non-rare species is independent of $S_\text{rare}$.  For $\alpha>1$ the contribution of the rare species decreases with $S_\text{rare}$ and vanishes for $S_\text{rare}\to\infty$.  Hence, the rare species contribute only weakly to the Hill diversity $D_\alpha$ for $\alpha>1$.  For $\alpha<1$ the contribution of the rare species increases with $S_\text{rare}$ and diverges for $S_\text{rare}\to\infty$.  Hence, for sufficiently large $S_\text{rare}$ the rare species contribution dominates the Hill diversity $D_\alpha$ for $\alpha<1$.  Note that the relative contribution of the rare to the non-rare species has a power-law dependence on $S_\text{rare}$ with exponent $1-\alpha$.  For the Hill diversity $D_1$ the relative contribution of the rare to the non-rare species has a logarithmic dependence on $S_\text{rare}$, see (\ref{eq:shannon}).

\newpage
\section*{Text S3} 
\subsection*{Hill diversities and rarefaction curve}
\bigskip

We follow \citet{Mao2007supp} to establish a link between the rarefaction curve $S_m$ and the Hill diversities $D_\alpha$.  Rewriting the sum $\sum_i p_i^\alpha$, we get
\begin{align*}
 \sum_{i=1}^S p_i^\alpha
 &= \sum_{i=1}^S \big( 1-(1-p_i) \big)^\alpha \\
 &= \sum_{i=1}^S \sum_{m=0}^\infty
    \frac{(-1)^m \Gamma(\alpha+1)}{m!\,\Gamma(\alpha-m+1)}
    (1-p_i)^m \\
 &= S \sum_{m=0}^\infty
    \frac{(-1)^m \Gamma(\alpha+1)}{m!\,\Gamma(\alpha-m+1)}
  - \sum_{m=0}^\infty
    \frac{(-1)^m \Gamma(\alpha+1)}{m!\,\Gamma(\alpha-m+1)}
    \sum_{i=1}^S \big( 1-(1-p_i)^m \big) \\
 &= \sum_{m=1}^\infty
    \frac{(-1)^{m+1} \Gamma(\alpha+1)}{m!\,\Gamma(\alpha-m+1)} S_m \\
 &= \sum_{m=1}^\infty
    \frac{\alpha\,\Gamma(m-\alpha)}{m!\,\Gamma(1-\alpha)} S_m
\end{align*}
where $\Gamma$ denotes the gamma function.  Hence,
\begin{equation}
 D_\alpha = \bigg( \sum_{m=1}^\infty
 \frac{\alpha\,\Gamma(m-\alpha)}{m!\,\Gamma(1-\alpha)}
 S_m \bigg)^{\frac{1}{1-\alpha}}.
 \label{eq:linkhill}
\end{equation}

We express the link with the rarefaction curve in terms of the Tsallis entropies $T_\alpha$ \citep{Tsallis1988supp},
\begin{equation*}
 T_\alpha = \frac{1}{1-\alpha} \Big( \sum_{i=1}^S p_i^\alpha -1 \Big),
\end{equation*}
which is closely related to the Hill diversities $D_\alpha$,
\begin{equation}
 D_\alpha = \big( 1 + (1-\alpha)T_\alpha \big)^\frac{1}{1-\alpha}.
 \label{eq:linktsallis}
\end{equation}
Equation~(\ref{eq:linkhill}) becomes
\begin{align*}
 T_\alpha
 &= \frac{1}{1-\alpha} \bigg( \alpha-1 + \sum_{m=2}^\infty
    \frac{\alpha\,\Gamma(m-\alpha)}{m!\,\Gamma(1-\alpha)}
    S_m \bigg) \\
 &= -1 + \sum_{m=2}^\infty
    \frac{\alpha\,\Gamma(m-\alpha)}{m!\,\Gamma(2-\alpha)} S_m.
\end{align*}
We study the behavior of the coefficients $c_m$ in this infinite sum,
\begin{equation*}
 c_m = \frac{\alpha\,\Gamma(m-\alpha)}{m!\,\Gamma(2-\alpha)}.
\end{equation*}
For $\alpha\in (0,2)$ all coefficients $c_m$ are positive, and
\begin{equation}
 c_m \sim m^{-(\alpha+1)} \qquad \qquad \text{as} \quad m \to\infty.
\label{eq:asymcffs}
\end{equation}
This shows that different Tsallis entropies $T_\alpha$ depend on different parts of the rarefaction curve $S_m$.  For $\alpha$ close to 2, the Tsallis entropy $T_\alpha$ is mainly determined by the rarefaction curve for small $m$.  For decreasing $\alpha$, the contribution of the rarefaction curve for large $m$ increases.  For the limit cases $\alpha\to 0$ and $\alpha\to 2$  the constant of proportionality in (\ref{eq:asymcffs}) vanishes.  For $\alpha=2$ we have $T_2 = 1-C = S_2-1$:  the only contribution of the rarefaction curve is at $m=2$.  For $\alpha=0$ we have $T_0 = S-1 = S_\infty-1$:  the contribution of the rarefaction curve is entirely shifted to $m\to\infty$.  This analysis also holds for the Hill diversities $D_\alpha$ because $D_\alpha$ is an increasing function of $T_\alpha$, see (\ref{eq:linktsallis}).

As an illustration, we apply (\ref{eq:linkhill}) to a community with a power-law tail.  That is, we consider an artificial community consisting of an infinite number of species, for which the species are arranged in decreasing order of abundance, and for which
\begin{equation*}
p_i \sim i^{-z} \qquad \qquad \text{as} \quad i \to\infty.
\end{equation*}
The abundances should be summable, so we have to impose that $z>1$.  The tail of the abundance distribution determines the asymptotic behavior of the rarefaction curve,
\begin{equation*}
S_m \sim m^{1/z} \qquad \qquad \text{as} \quad m \to\infty.
\end{equation*}
From (\ref{eq:linkhill}) and (\ref{eq:asymcffs}) it follows that the diversity $D_\alpha$ is finite for $\alpha > \frac{1}{z}$, and diverges for $\alpha \leq \frac{1}{z}$.  This can be checked directly from Definition~(\ref{eq:defhill}).

\newpage
\section*{Text S4} 
\subsection*{Estimating species abundances from sample data}
\bigskip

The Good-Turing estimators \citep{Good1953supp} are a well-known family of frequency estimators.  Here we present a compact derivation, given in \citet{Nadas1985supp}, which demonstrates that the Good-Turing estimators are non-parametric, that is, free of assumptions about the abundance distribution.

Let $\Theta$ be a random variable taking values between 0 and 1, with a distribution function $G(\theta)$ about which nothing is known.  Suppose that $R$ is another random variable whose conditional distribution $p_M(r | \theta)$, when $\Theta$ has the value $\theta$, is binomial with parameters $M$ and $\theta$,
\begin{equation}
p_M(r | \theta) = \binom{M}{r} \theta^r (1-\theta)^{M-r}.
\end{equation}
Then we have the identity
\begin{equation}
\theta \; p_M(r | \theta) = \frac{r+1}{M+1} \; p_{M+1}(r+1 | \theta)
\label{eq:nadas1}
\end{equation}
Suppose now that we wish to estimate the value of $\theta$ given that $R$ is observed to take the value $r$.  Taking a Bayesian approach with prior distribution $G$, the posterior mean for $\theta$ is
\begin{equation}
E[\theta | R=r] = \frac{r+1}{M+1} \, \frac{p_{M+1}(r+1)}{p_M(r)}
\label{eq:nadas2}
\end{equation}
where $p_M$ is the unconditional probability mass function of $R$ (that is, integrated out over $G$).  This derivation is non-parametric in that $G$ is not only unknown, but no assumptions are made about $G$: the probability mass function $p_M$ must therefore be estimated directly from the sample data, so that we are in fact performing empirical Bayes estimation.

In the context of diversity estimation, we regard $G$ as the community abundance distribution, $\theta$ as the species abundance to be estimated and $r$ as the number of times that this species occurs in the sample.  We use the maximum likelihood estimates for $p_M(r)$ and $p_{M+1}(r+1)$ given by $F_r/M$ and $F_{r+1}/(M+1)$, respectively.  Plugging the estimates into (\ref{eq:nadas2}) and assuming that $M \gg 1$, we get the estimated community abundance $\widehat \theta_r$ of a species observed $r$ times in the sample,
\begin{equation}
\widehat{\theta}_r = \frac{r+1}{M} \, \frac{F_{r+1}}{F_r},
\label{eq:nadas3}
\end{equation}
which are the Good-Turing frequency estimators.

As a corollary of (\ref{eq:nadas3}) we get the estimator for the total abundance of the observed species,
\begin{equation*}
\sum_{r\geq 1} F_r \; \widehat \theta_r
= \frac{1}{M} \sum_{r\geq 1} (r+1) F_{r+1}
= \frac{M - F_1}{M},
\end{equation*}
so that the total abundance $p_\text{unobs}$ of the unobserved species is estimated as
\begin{equation}
\widehat p_\text{unobs} = \frac{F_1}{M}.
\label{eq:nadas4}
\end{equation}
In words, the total relative abundance of unobserved species in the community is estimated as the total relative abundance of singletons in the sample.

\newpage
\section*{Text S5} 
\subsection*{Estimating Hill diversities from sample data}
\bigskip

We construct estimators for the Hill diversity $D_\alpha$ based on a sample of size $M$ taken from the community.  Our strategy consists in first estimating the rarefaction curve $S_m$ and then using the link (\ref{eq:linkhill}) between $D_\alpha$ and $S_m$.

The estimation of the rarefaction curve decomposes into two parts.  For the part $m\leq M$ the rarefaction curve can be estimated unbiasedly using the estimator (\ref{eq:rfactestim}).  For the part $m>M$ the sample data have to be extrapolated, and no unbiased estimator exists.  We denote the relative abundances of the unobserved species by $q_1, q_2, \ldots$ (there are $S - S_\text{obs}$ unobserved species).  If we knew the abundances $q_i$, then we could compute the rarefaction curve using the formula,
\begin{equation}
\widehat S_m = S_\text{obs} + \sum_{i\geq 1}
\Big( 1-\big(1-q_i\big)^{m-M} \Big)
\qquad\qquad m > M.
\label{eq:rfactestimunobs}
\end{equation}
As we have argued in the main text, the sample data contain little information about the abundances $q_i$ of unobserved species. However, the Good-Turing estimator~(\ref{eq:nadas4}) for the total abundance $p_\text{unobs}= \sum_{i\geq 1} q_i $ of the unobserved species is available.  It follows from (\ref{eq:rfactestimunobs}) that the estimation of the rarefaction curve $S_m$ for $m>M$ reduces to distributing the estimated abundance $\widehat p_\text{unobs}$ over the individual unobserved species.

We work out two scenarios, see Figure~3 of the main text.  In the first scenario we distribute $\widehat p_\text{unobs}$ so as to obtain the lowest possible value of the diversity $D_\alpha$ \emph{consistent with the sample data}.  By this we mean that $\widehat p_\text{unobs}$ must be distributed in a manner which remains consistent with the estimates $\widehat \theta_r$.  The lowest diversity occurs when all unobserved species have the same abundance, $q_1=q_2=\ldots\;=q^-$, and this abundance is as high as possible.  However, as noted in \citet{Good1953supp}, the frequency estimates $\widehat \theta_r$ must increase as $r$ increases: this implies an upper bound for $q^-$, namely $\widehat \theta_1$ (which is the estimated community abundance of any species observed exactly once in the sample).  We therefore take $q^- = \widehat \theta_1 = \frac{2 F_2}{M F_1}$ so that, from (\ref{eq:nadas4}), there are $\frac{F_1^2}{2 F_2}$ unobserved species.  Hence, the estimated rarefaction curve~(\ref{eq:rfactestimunobs}) becomes
\begin{equation}
\widehat S_m^- = S_\text{obs} + \frac{F_1^2}{2 F_2}
\bigg( 1 - \Big( 1 - \frac{2 F_2}{M F_1} \Big)^{m-M} \bigg)
\qquad\qquad m > M,
\label{eq:rfactestimlower}
\end{equation}
where the superscript in $\widehat S_m^-$ indicates the low-diversity scenario.

In the second scenario we distribute $\widehat p_\text{unobs}$ so as to obtain the highest possible value of the diversity $D_\alpha$.  The highest diversity is obtained when all unobserved species have the same abundance, $q_1 = q_2 = \ldots = q^+$, and this abundance is as small as possible.  The smallest abundance a species can have in a community of size $N$ is equal to $\frac{1}{N}$, corresponding to a species represented by a single individual.  We therefore take $q^+ = \frac{1}{N}$ so that, from (\ref{eq:nadas4}), there are $\frac{N F_1}{M}$ unobserved species.  Hence, the estimated rarefaction curve~(\ref{eq:rfactestimunobs}) becomes
\begin{equation}
\widehat S_m^+ = S_\text{obs} + \frac{N F_1}{M}
\bigg( 1 - \Big( 1 - \frac{1}{N} \Big)^{m-M} \bigg)
\qquad\qquad m > M,
\label{eq:rfactestimupper}
\end{equation}
where the superscript in $\widehat S_m^+$ indicates the high-diversity scenario.  Note that the upper estimator (\ref{eq:rfactestimupper}) depends on the community size $N$, in contrast to the estimator (\ref{eq:rfactestimlower}).

To summarize, we have obtained two estimators for the Hill diversity $D_\alpha$, a lower estimate $\widehat D_\alpha^-$ and an upper estimate $\widehat D_\alpha^+$.  They can be computed as follows:
\begin{description}
\item[Lower estimate] First, compute the lower estimate of the rarefaction curve.  From (\ref{eq:rfactestim}) and (\ref{eq:rfactestimlower}),
\begin{equation}
 \widehat S_m^- = \begin{cases}
 \sum_{k \geq 1} F_k \Big( 1 - \frac{\binom{M-k}{m}}{\binom{M}{m}} \Big)
 & \text{if $m=1,2,\ldots,M$} \\
 S_\text{obs} + \frac{F_1^2}{2 F_2}
 \Big( 1 - \big( 1 - \frac{2 F_2}{M F_1} \big)^{m-M} \Big)
 & \text{if $m=M+1,M+2,\ldots$}
 \end{cases}
\end{equation}
Then, substitute this result into (\ref{eq:linkhill}) to estimate the Hill diversity,
\begin{equation}
 \widehat D_\alpha^- = \bigg( \sum_{m=1}^\infty
 \frac{\alpha\;\Gamma(m-\alpha)}{m!\;\Gamma(1-\alpha)}\,
 \widehat S_m^- \bigg)%
 ^{\frac{1}{1-\alpha}}.
\end{equation}
\item[Upper estimate] First, compute the upper estimate of the rarefaction curve.  From (\ref{eq:rfactestim}) and (\ref{eq:rfactestimupper}),
\begin{equation}
 \widehat S_m^+ = \begin{cases}
 \sum_{k \geq 1} F_k \Big( 1 - \frac{\binom{M-k}{m}}{\binom{M}{m}} \Big)
 & \text{if $m=1,2,\ldots,M$} \\
 S_\text{obs} + \frac{N F_1}{M}
 \Big( 1 - \big( 1 - \frac{1}{N} \big)^{m-M} \Big)
 & \text{if $M+1,M+2,\ldots$}
 \end{cases}
\end{equation}
Then, substitute this result into (\ref{eq:linkhill}) to estimate the Hill diveristy,
\begin{equation}
 \widehat D_\alpha^+ = \bigg( \sum_{m=1}^\infty
 \frac{\alpha\;\Gamma(m-\alpha)}{m!\;\Gamma(1-\alpha)}\,
 \widehat S_m^+ \bigg)%
 ^{\frac{1}{1-\alpha}}.
\end{equation}
\end{description}
The Matlab code to compute the Hill diversity estimates $\widehat D_\alpha^-$ and $\widehat D_\alpha^+$ is part of the Supplementary Information.

We discuss three properties of the estimators $\widehat D_\alpha^-$ and $\widehat D_\alpha^+$ that follow directly from their definitions.  First, the lower estimate $\widehat D_\alpha^-$ generalizes Chao's estimator for species richness,
\begin{equation*}
\widehat D_0^- = \widehat S_\infty^- = S_\text{obs} + \frac{F_1^2}{2F_2}.
\end{equation*}
Note that the lower estimate, like Chao's estimator, only gives meaningful results if the number of species observed once or twice in the sample is sufficiently large, and at least $F_2 > 0$.  These conditions are typically satisfied in practice, especially for highly diverse communities.

Second, the upper estimate $\widehat D_\alpha^+$ depends on community size $N$, which is typically several orders of magnitude larger than sample size $M$.  It is therefore instructive to consider the limit $N\to\infty$.  A computation analogous to the one in Text~S2 shows that the upper estimate $\widehat D_\alpha^+$ diverges as $N^{1-\alpha}$ for $\alpha < 1$, and as $\log N$ for $\alpha=1$.  Hence, we expect large values of the upper estimate (and therefore large estimation uncertainty) for $\alpha<1$, especially for $\alpha$ close to zero (that is, close to species richness).

Third, the estimators $\widehat D_\alpha^-$ and $\widehat D_\alpha^+$ coincide for the Simpson diversity.  The Simpson diversity $D_2$ is the only Hill diversity $D_\alpha$ that does not depend on the extrapolation of the rarefaction curve.  It is a function of the rarefaction curve at $m=2$:  $D_2 = \frac{1}{2-S_2}$.  Because the initial part of the estimated rarefaction curve is the same for the lower and upper estimate, the Simpson diversity estimates are equal, $\widehat D_2^- = \widehat D_2^+$.  The Simpson diversity is not sensitive to the extrapolation of the rarefaction curve, and therefore easy to estimate.

\newpage

\noindent {\bf\LARGE Supplementary Tables}

\section*{Table S1} 

\begin{table}[!ht]
\caption{Description of communities used in Figure~2.  Communities C1, C2 and C3 have a power-law abundance distribution, with parameters $S$, the number of species in the community, and $z$, the exponent of the power-law.  The Hill diversity of order $\alpha=0$ is equal to the number of species, $D_0 = S$;  the Hill diversity of order $\alpha=1$ is the Shannon diversity;  the Hill diversity of order $\alpha=2$ is the Simpson diversity.  For a sample of size $2\;10^4$, the number of observed species is denoted by $S_\text{obs}$ and Chao's estimator for species richness is denoted by $\widehat S_\text{Chao}$.\label{tab:descrcomm}}
\begin{center}
\begin{tabular}{l|cccccc}
& $S$ & $z$ & $D_1$ & $D_2$ & $S_\text{obs}$ & $\widehat S_\text{Chao}$ \\
\hline
community C1 & $5\;10^4$ & $1.1$ & $640$ & $35$ & $4.8\;10^3$ & $1.5\;10^4$ \\
community C2 & $2\;10^5$ & $1.3$ & $100$ & $11$ & $2.4\;10^3$ & $8.3\;10^3$ \\
community C3 & $10^6$ & $1.6$ & $15$ & $4.5$ & $690$ & $1.8\;10^3$
\end{tabular}
\end{center}
\end{table}

\clearpage\newpage

\section*{Table S2} 

\begin{table}[!ht]
\caption{Data for empirically-sampled microbial communities.  We report the sample size $M$, the number of species observed in the sample $S_\text{obs}$, the number of singleton species $F_1$, that is, the number of species that have been sampled only once, the estimated relative abundance of the unobserved species $\widehat p_\text{unobs}$, and the Chao estimate $\widehat S_\text{Chao}$ for the number of species in the community.  The data sets are taken from \citet{Quince2008supp}: a seawater bacterial sample from the upper ocean \citep{Rusch2007supp}, soil bacterial samples at four locations: Brazil, Florida, Illinois and Canada \citep{Roesch2007supp}, and seawater samples from deep-sea vents at two locations: FS312 and FS396, separated into bacteria and archaea \citep{Huber2007supp}.\label{tab:empirical}}
\begin{center}
\begin{tabular}{l|cccccc}
& $M$ & $S_\text{obs}$ & $F_1$
 & $\widehat p_\text{unobs}$ & $\widehat S_\text{Chao}$ \\
\hline
upper ocean & $7068$ & $811$ & $311$ & $0.044$ & $1038$ \\
soil, Brazil & $26079$ & $2880$ & $1176$ & $0.045$ & $4604$ \\
soil, Florida & $28150$ & $3440$ & $1541$ & $0.055$ & $5643$ \\
soil, Illinois & $31621$ & $3357$ & $1466$ & $0.046$ & $5745$ \\
soil, Canada & $52773$ & $5515$ & $2634$ & $0.050$ & $10394$ \\
FS312, bacteria & $442062$ & $12183$ & $5339$ & $0.012$ & $19568$ \\
FS312, archaea & $200199$ & $1594$ & $460$ & $0.002$ & $2175$ \\
FS396, bacteria & $247826$ & $5843$ & $2825$ & $0.011$ & $10570$ \\
FS396, archaea & $16428$ & $418$ & $158$ & $0.010$ & $630$
\end{tabular}
\end{center}
\end{table}

\clearpage\newpage

\noindent {\bf\LARGE Supplementary Figures}

\section*{Figure S1} 

\clearpage\newpage

\section*{Figure S2} 

\begin{figure}[!ht]
\begin{center}
\includegraphics[width=.6\textwidth]{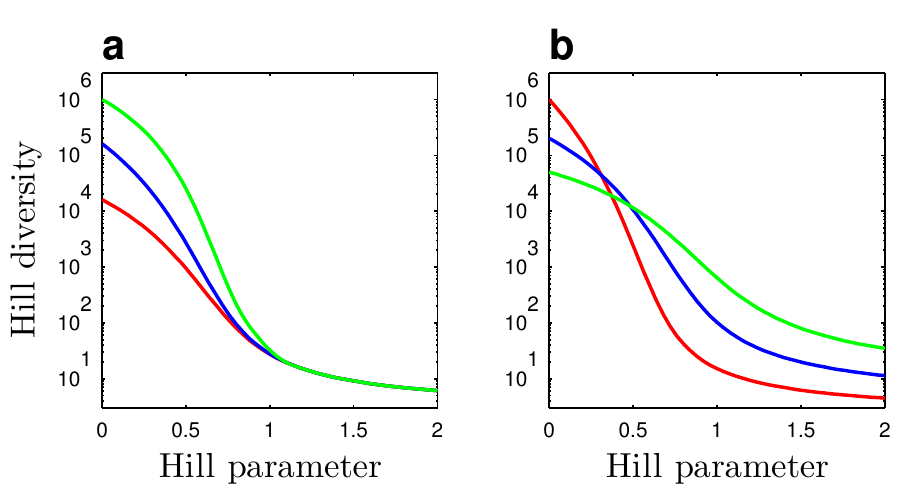}
\end{center}
\caption{Hill diversity for large $\alpha$ is insensitive to rare species tail.  Panel~a:  We computed the Hill diversity $D_\alpha$ for the three communities of Figure~\ref{fig:figS1}.  The Hill diversities for $\alpha > 1$ almost coincide because the communities have the same set of non-rare species.  The Hill diversities for $\alpha < 1$ differ because the communities have different rare species tails.  Panel~b:  We computed the Hill diversity $D_\alpha$ for the three communities of Figure~2.  The curves of Hill diversities intersect.  For small $\alpha$, the most species-rich community (C3, green) has the largest Hill diversity, and the most species-poor community (C1, red) has the smallest Hill diversity.  For larger $\alpha$, the most even community (C1, red) has the largest Hill diversity, and the most uneven community (C3, green) has the smallest Hill diversity.\label{fig:figS2}}
\end{figure}

\clearpage\newpage

\section*{Figure S3} 

\begin{figure}[!ht]
\begin{center}\includegraphics[width=4.8in]{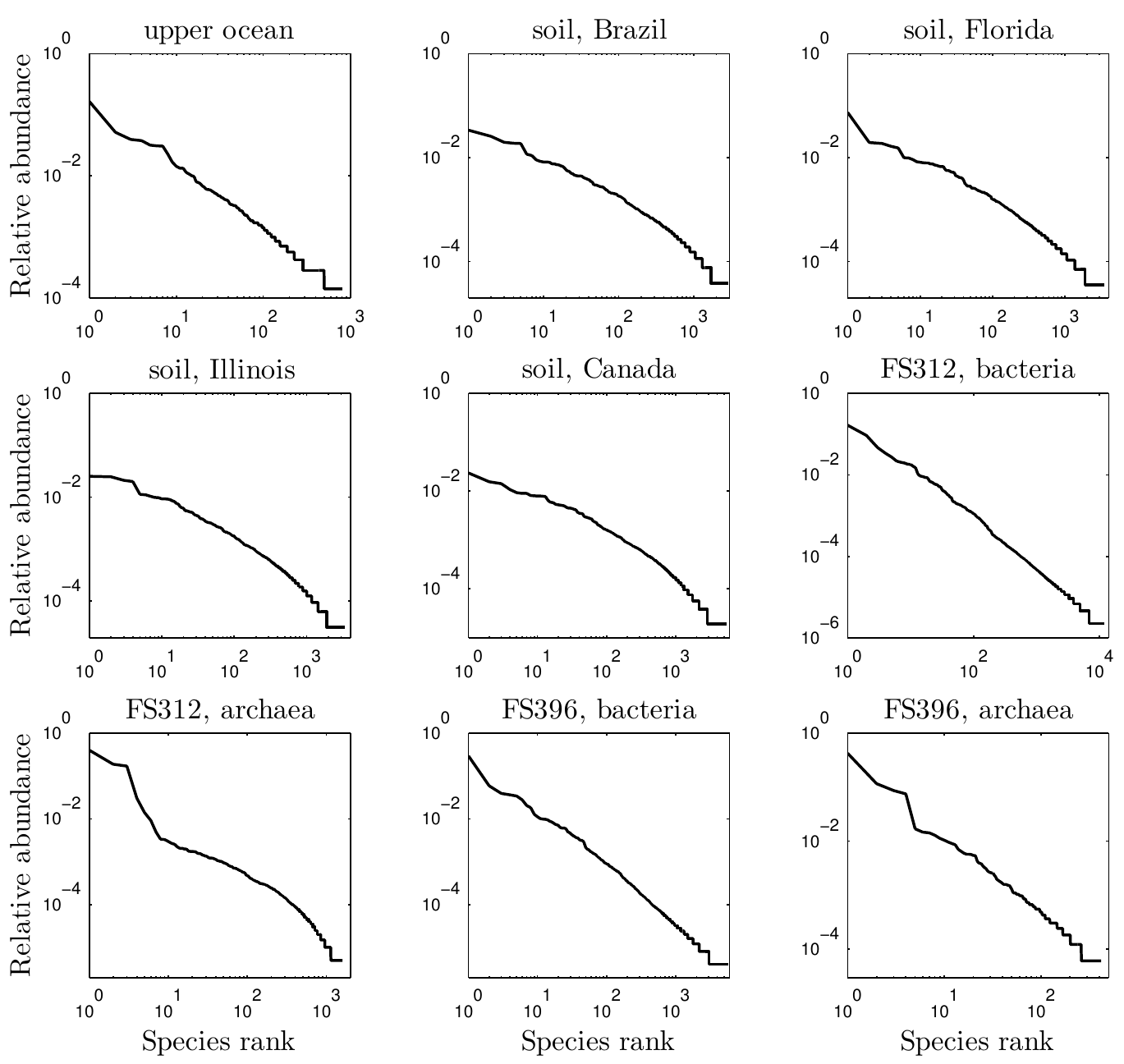}
\end{center}
\caption{Rank-abundances curve of empirical microbial community samples.  Relative abundance in the sample is plotted against species rank in the sample.  We used the same data sets as \citet{Quince2008supp}: a seawater bacterial sample from the upper ocean \citep{Rusch2007supp}, soil bacterial samples at four locations: Brazil, Florida, Illinois and Canada \citep{Roesch2007supp}, and seawater samples from deep-sea vents at two locations: FS312 and FS396, separated into bacteria and archaea \citep{Huber2007supp}.\label{fig:figS3}}
\end{figure}

\clearpage\newpage

\section*{Figure S4} 

\begin{figure}[!ht]
\begin{center}\includegraphics[width=5in]{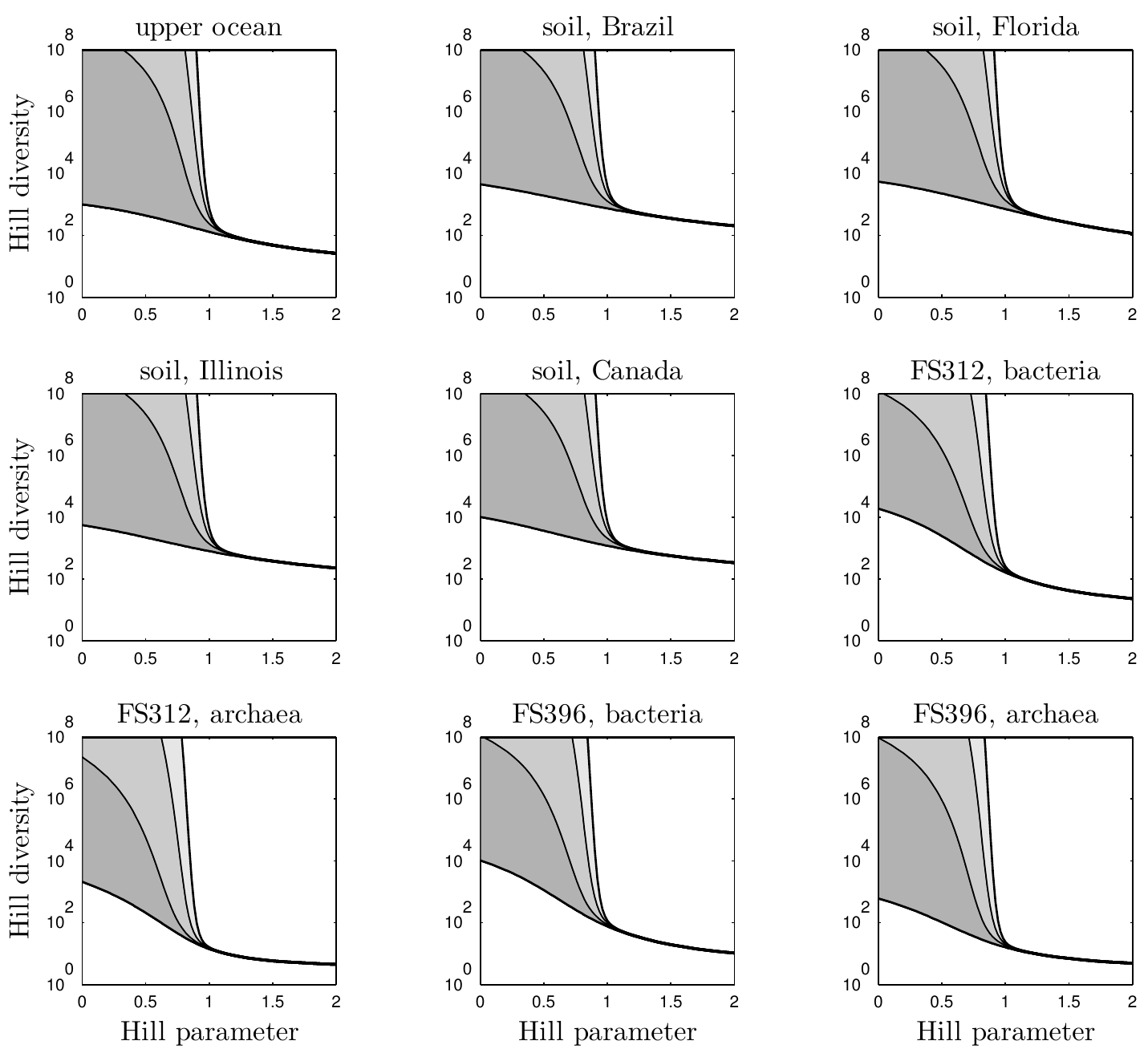}
\end{center}
\caption{Community-size dependence of Hill diversity estimates.  Same data sets as in Figure~5, but for three values of community size $N$.  The lower estimate is independent of $N$;  the upper estimate increases with increasing $N$  (from left to right: $N=10^{10}$, $N=10^{15}$, $N=10^{20}$).  We observe the same behavior as for the \emph{in silico} generated data sets of Figure~4.\label{fig:figS4}}
\end{figure}

\clearpage\newpage



\begin{thebibliography}{55}

\bibitem[Bent and Forney(2008)]{Bent2008}
Bent SJ, Forney LJ. (2008).
\newblock The tragedy of the uncommon: understanding limitations in the
  analysis of microbial diversity.
\newblock \emph{ISME J} \textbf{2}: 689--695.

\bibitem[Bohannan and Hughes(2003)]{Bohannan2003}
Bohannan BJM, Hughes JB. (2003).
\newblock New approaches to analyzing microbial biodiversity data.
\newblock \emph{Curr Opin Microbiol} \textbf{6}: 282--287.

\bibitem[Brose \emph{et~al.}(2003)Brose, Martinez, and Williams]{Brose2003}
Brose U, Martinez ND, Williams RJ. (2003).
\newblock Estimating species richness: Sensitivity to sample coverage and
  insensitivity to spatial patterns.
\newblock \emph{Ecology} \textbf{84}: 2364--2377.

\bibitem[Bunge(2009)]{Bunge2009}
Bunge J. (2009).
\newblock Statistical estimation of uncultivated microbial diversity.
\newblock In:  Epstein SS (ed). \emph{Uncultivated Microorganisms},
Springer-Verlag, pp 1--18.

\bibitem[Bunge and Fitzpatrick(1993)]{Bunge1993}
Bunge J, Fitzpatrick M. (1993).
\newblock Estimating the number of species: A review.
\newblock \emph{J Amer Statist Assoc} \textbf{88}: 364--373.

\bibitem[Chao(1984)]{Chao1984}
Chao A. (1984).
\newblock Nonparametric estimation of the number of classes in a population.
\newblock \emph{Scand J Statist} \textbf{11}: 265--270.

\bibitem[Chao \emph{et~al.}(2009)Chao, Colwell, Lin, and Gotelli]{Chao2009}
Chao A, Colwell RK, Lin CW, Gotelli NJ. (2009).
\newblock Sufficient sampling for asymptotic minimum species richness
  estimators.
\newblock \emph{Ecology} \textbf{90}: 1125--1133.

\bibitem[Colwell \emph{et~al.}(2004)Colwell, Mao, and Chang]{Colwell2004}
Colwell RK, Mao CX, Chang J. (2004).
\newblock Interpolating, extrapolating, and comparing incidence-based species
  accumulation curves.
\newblock \emph{Ecology} \textbf{85}: 2717--2727.

\bibitem[Curtis \emph{et~al.}(2002)Curtis, Sloan, and Scannell]{Curtis2002}
Curtis TP, Sloan WT, Scannell JW.  (2002).
\newblock Estimating prokaryotic diversity and its limits.
\newblock \emph{Proc Natl Acad Sci USA} \textbf{99}: 10494--10499.

\bibitem[Dykhuizen(1998)]{Dykhuizen1998}
Dykhuizen DE. (1998).
\newblock Santa {R}osalia revisited: why are there so many species of bacteria?
\newblock \emph{Antonie Van Leeuwenhoek} \textbf{73}: 25--33.

\bibitem[Engen(1978)]{Engen1978}
Engen S. (1978).
\newblock \emph{Stochastic Abundance Models}.
\newblock Chapman \& Hall.

\bibitem[Fromin \emph{et~al.}(2002)Fromin, Hamelin, Tarnawski, Roesti,
  Jourdain-Miserez, Forestier, Teyssier-Cuvelle, Gillet, Aragno, and
  Rossi]{Fromin2002}
Fromin N, Hamelin J, Tarnawski S, Roesti D, Jourdain-Miserez K,
  Forestier N, \emph{et al.} (2002).
\newblock Statistical analysis of denaturing gel electrophoresis ({DGE})
  fingerprinting patterns.
\newblock \emph{Environ Microbiol} \textbf{4}: 634--643.

\bibitem[Gans \emph{et~al.}(2005)Gans, Wolinsky, and Dunbar]{Gans2005}
Gans J, Wolinsky M, Dunbar J. (2005).
\newblock Computational improvements reveal great bacterial diversity and high
  metal toxicity in soil.
\newblock \emph{Science} \textbf{309}: 1387--1390.

\bibitem[Gobet \emph{et~al.}(2010)Gobet, Quince, and Ramette]{Gobet2010}
Gobet A, Quince C, Ramette A. (2010).
\newblock Multivariate cutoff level analysis ({MultiCoLA}) of large community
  data sets.
\newblock \emph{Nucl Acids Res} \textbf{38}: e155.

\bibitem[Gotelli and Colwell(2001)]{Gotelli2001}
Gotelli NJ, Colwell RK. (2001).
\newblock Quantifying biodiversity: Procedures and pitfalls in the measurement
  and comparison of species richness.
\newblock \emph{Ecol Lett} \textbf{4}: 379--391.

\bibitem[Gotelli and Colwell(2011)]{Gotelli2011}
Gotelli NJ, Colwell RK. (2011).
\newblock Estimating species richness.
\newblock In: Magurran AE, McGill BJ (eds). \emph{Biological
  Diversity: Frontiers in Measurement and Assessment}.
  Oxford University Press, pp 39--54.

\bibitem[Green \emph{et~al.}(2008)Green, Bohannan, and Whitaker]{Green2008}
Green JL, Bohannan BJM, Whitaker RJ. (2008).
\newblock Microbial biogeography: From taxonomy to traits.
\newblock \emph{Science} \textbf{320}: 1039--1043.

\bibitem[Haegeman \emph{et~al.}(2008)Haegeman, Vanpeteghem, Godon, and
  Hamelin]{Haegeman2008}
Haegeman B, Vanpeteghem D, Godon JJ, Hamelin J. (2008).
\newblock {DNA} reassociation kinetics and diversity indices: richness is not
  rich enough.
\newblock \emph{Oikos} \textbf{117}: 177--181.

\bibitem[Hill(1973)]{Hill1973}
Hill MO. (1973).
\newblock Diversity and evenness: A unifying notation and its consequences.
\newblock \emph{Ecology} \textbf{54}: 427--432.

\bibitem[Hong \emph{et~al.}(2006)Hong, Bunge, Jeon, and Epstein]{Hong2006}
Hong SH, Bunge J, Jeon SO, Epstein SS. (2006).
\newblock Predicting microbial species richness.
\newblock \emph{Proc Natl Acad Sci USA} \textbf{103}: 117--122.

\bibitem[Horner-Devine and Bohannan(2006)]{HornerDevine2006}
Horner-Devine MC, Bohannan BJM. (2006).
\newblock Phylogenetic clustering and overdispersion in bacterial communities.
\newblock \emph{Ecology} \textbf{87}: S100--S108.

\bibitem[Huber \emph{et~al.}(2007)Huber, Welch, Morrison, Huse, Neal, Butterfield, and
  Sogin]{Huber2007}
Huber JA, Welch DBM, Morrison HG, Huse SM, Neal PR, Butterfield DA
\emph{et al.} (2007).
\newblock Microbial population structures in the deep marine biosphere.
\newblock \emph{Science} \textbf{318}: 97--100.

\bibitem[Hughes \emph{et~al.}(2001)Hughes, Hellmann, Ricketts, and
  Bohannan]{Hughes2001}
Hughes JB, Hellmann JJ, Ricketts TH, Bohannan BJM. (2001).
\newblock Counting the uncountable: statistical approaches to estimating
  microbial diversity.
\newblock \emph{Appl Environ Microbiol} \textbf{67}: 4399--4406.

\bibitem[Ives and Carpenter(2007)]{Ives2007}
Ives A, Carpenter S. (2007).
\newblock Stability and diversity of ecosystems.
\newblock \emph{Science} \textbf{317}: 58--68.

\bibitem[Jost(2006)]{Jost2006}
Jost L. (2006).
\newblock Entropy and diversity.
\newblock \emph{Oikos} \textbf{113}: 363--375.

\bibitem[Kemp and Aller(2004)]{Kemp2004}
Kemp P, Aller J. (2004).
\newblock Bacterial diversity in aquatic and other environments: what {16S}
  {rDNA} libraries can tell us.
\newblock \emph{FEMS Microbiol Ecol} \textbf{47}: 161--171.

\bibitem[Kislyuk \emph{et~al.}(2011)Kislyuk, Haegeman, Bergman, and
  Weitz]{Kislyuk2011}
Kislyuk AO, Haegeman B, Bergman NH, Weitz JS. (2011).
\newblock Genomic fluidity: an integrative view of gene diversity within
  microbial populations.
\newblock \emph{BMC Genomics} \textbf{12}: 32.

\bibitem[Lande \emph{et~al.}(2000)Lande, DeVries, and Walla]{Lande2000}
Lande R, DeVries PJ, Walla TR. (2000).
\newblock When species accumulation curves intersect: implications for ranking
  diversity using small samples.
\newblock \emph{Oikos} \textbf{89}: 601--605.

\bibitem[Loisel \emph{et~al.}(2006)Loisel, Harmand, Zemb, Latrille, Lobry, Delgen\`es,
  and Godon]{Loisel2006}
Loisel P, Harmand J, Zemb O, Latrille E, Lobry C, Delgen\`es JP, \emph{et al.} (2006).
\newblock Denaturing gradient electrophoresis ({DGE}) and single-strand
  conformation polymorphism ({SSCP}) molecular fingerprintings revisited by
  simulation and used as a tool to measure microbial diversity.
\newblock \emph{Environ Microbiol} \textbf{8}: 720--731.

\bibitem[Loreau(2010)]{Loreau2010}
Loreau M. (2010).
\newblock \emph{From Populations to Ecosystems: Theoretical Foundations for a
  New Ecological Synthesis}.
\newblock Princeton University Press.

\bibitem[Loreau \emph{et~al.}(2001)Loreau, Naeem, Inchausti, Bengtsson, Grime, Hector,
  Hooper, Huston, Raffaelli, Schmid, Tilman, and Wardle]{Loreau2001}
Loreau M, Naeem S, Inchausti P, Bengtsson J, Grime JP, Hector A, \emph{et al.} (2001).
\newblock Biodiversity and ecosystem functioning: Current knowledge and future
  challenges.
\newblock \emph{Science} \textbf{294}: 804--808.

\bibitem[Lozupone and Knight(2007)]{Lozupone2007}
Lozupone CA, Knight R. (2007).
\newblock Global patterns in bacterial diversity.
\newblock \emph{Proc Natl Acad Sci USA} \textbf{104}: 11436--11440.

\bibitem[Magurran(2004)]{Magurran2004}
Magurran AE. (2004).
\newblock \emph{Measuring Biological Diversity}.
\newblock Blackwell Publishing.

\bibitem[Mao and Colwell(2005)]{Mao2005}
Mao CX, Colwell RK. (2005).
\newblock Estimation of species richness: Mixture models, the role of rare
  species, and inferential challenges.
\newblock \emph{Ecology} \textbf{86}: 1143--1153.

\bibitem[May(1988)]{May1988}
May RM. (1988).
\newblock How many species are there on earth?
\newblock \emph{Science} \textbf{241}: 1441--1449.

\bibitem[Mora \emph{et~al.}(2011)Mora, Tittensor, Adl, Simpson, and Worm]{Mora2011}
Mora C, Tittensor DP, Adl S, Simpson AGB, Worm B. (2011).
\newblock How many species are there on earth and in the ocean?
\newblock \emph{PLoS Biol} \textbf{9}: e1001127.

\bibitem[{\O}vre{\aa}s and Curtis(2011)]{Ovreas2011}
{\O}vre{\aa}s L, Curtis TP. (2011).
\newblock Microbial diversity and ecology.
\newblock In: Magurran AE, McGill BJ (eds). \emph{Biological
  Diversity: Frontiers in Measurement and Assessment}.
  Oxford University Press, pp 221--236.

\bibitem[Pedr\'os-Ali\'o(2006)]{PedrosAlio2006}
Pedr\'os-Ali\'o C. (2006).
\newblock Marine microbial diversity: can it be determined?
\newblock \emph{Trends Microbiol} \textbf{14}: 257--263.

\bibitem[Pedr\'os-Ali\'o(2007)]{PedrosAlio2007}
Pedr\'os-Ali\'o C. (2007).
\newblock Dipping into the rare biosphere.
\newblock \emph{Science} \textbf{315}: 192--193.

\bibitem[Quince \emph{et~al.}(2008)Quince, Curtis, and Sloan]{Quince2008}
Quince C, Curtis TP, Sloan WT. (2008).
\newblock The rational exploration of microbial diversity.
\newblock \emph{ISME J} \textbf{2}: 997--1006.

\bibitem[Roesch \emph{et~al.}(2007)Roesch, Fulthorpe, Riva, Casella, Hadwin, Kent,
  Daroub, Camargo, Farmerie, and Triplett]{Roesch2007}
Roesch LFW, Fulthorpe RR, Riva A, Casella G, Hadwin AKM, Kent AD
\emph{et al.} (2007).
\newblock Pyrosequencing enumerates and contrasts soil microbial diversity.
\newblock \emph{ISME J} \textbf{1}: 283---290.

\bibitem[Rusch \emph{et~al.}(2007)Rusch, Halpern, Sutton, Heidelberg, Williamson,
  Yooseph, Wu, Eisen, Hoffman, Remington, Beeson, Tran, Smith, Baden-Tillson,
  Stewart, Thorpe, Freeman, Andrews-Pfannkoch, Venter, Li, Kravitz, Heidelberg,
  Utterback, Rogers, Falc\'on, Souza, Bonilla-Rosso, Eguiarte, Karl,
  Sathyendranath, Platt, Bermingham, Gallardo, Tamayo-Castillo, Ferrari,
  Strausberg, Nealson, Friedman, Frazier, and Venter]{Rusch2007}
Rusch DB, Halpern AL, Sutton G, Heidelberg KB, Williamson S, Yooseph S
\emph{et al.} (2007).
\newblock The \emph{Sorcerer II} {G}lobal {O}cean {S}ampling expedition:
  {N}orthwest {A}tlantic through {E}astern {T}ropical {P}acific.
\newblock \emph{PLoS Biol} \textbf{5}: e77.

\bibitem[Schloss and Handelsman(2005)]{Schloss2005}
Schloss PD, Handelsman J. (2005).
\newblock Introducing {DOTUR}, a computer program for defining operational
  taxonomic units and estimating species richness.
\newblock \emph{Appl Environ Microbiol} \textbf{71}: 1501--1506.

\bibitem[Schloss and Handelsman(2006)]{Schloss2006}
Schloss PD, Handelsman J. (2006).
\newblock Toward a census of bacteria in soil.
\newblock \emph{PLoS Comput Biol} \textbf{2}: e92.

\bibitem[Shannon(1948)]{Shannon1948}
Shannon CE. (1948).
\newblock A mathematical theory of communication.
\newblock \emph{Bell System Tech J}, \textbf{27}: 379--423 and 623--656.

\bibitem[Shaw \emph{et~al.}(2008)Shaw, Halpern, Beeson, Tran, Venter, and
  Martiny]{Shaw2008}
Shaw AK, Halpern AL, Beeson K, Tran B, Venter JC, Martiny JBH. (2008).
\newblock It's all relative: ranking the diversity of aquatic bacterial
  communities.
\newblock \emph{Environ Microbiol} \textbf{10}: 2200--2210.

\bibitem[Shen \emph{et~al.}(2003)Shen, Chao, and Lin]{Shen2003}
Shen TJ, Chao A, Lin CF. (2003).
\newblock Predicting the number of new species in further taxonomic sampling.
\newblock \emph{Ecology} \textbf{84}: 798--804.

\bibitem[Simpson(1949)]{Simpson1949}
Simpson EH. (1949).
\newblock Measurement of diversity.
\newblock \emph{Nature} \textbf{163}: 688.

\bibitem[Sloan \emph{et~al.}(2008)Sloan, Quince, and Curtis]{Sloan2008}
Sloan WT, Quince C, Curtis TP. (2008).
\newblock The uncountables.
\newblock In: Zengler K (ed). \emph{Accessing Uncultivated Microorganisms:
  From the Environment to Organisms and Genomes and Back}. ASM
  Press, pp 35--54.

\bibitem[Sogin \emph{et~al.}(2006)Sogin, Morrison, Huber, Welch, Huse, Neal, Arrieta,
  and Herndl]{Sogin2006}
Sogin ML, Morrison HG, Huber JA, Welch DM, Huse SM, Neal PR,
\emph{et al.} (2006).
\newblock Microbial diversity in the deep sea and the underexplored ``rare
  biosphere''.
\newblock \emph{Proc Natl Acad Sci USA} \textbf{103}: 12115--12120.

\bibitem[Stackebrandt \emph{et~al.}(2002)Stackebrandt, Frederiksen, Garrity, Grimont,
  K{\"a}mpfer, Maiden, Nesme, Rossello-Mora, Swings, Tr{\"u}per, Vauterin,
  Ward, and Whitman]{Stackebrandt2002}
Stackebrandt E, Frederiksen W, Garrity GM, Grimont PAD, K{\"a}mpfer P, Maiden MCJ,
\emph{et al.} (2002).
\newblock Report of the ad hoc committee for the re-evaluation of the species
  definition in bacteriology.
\newblock \emph{Int J Syst Evol Microbiol} \textbf{52}: 1043--1047.

\bibitem[Tettelin \emph{et~al.}(2005)Tettelin, Masignani, Cieslewicz, Donati, Medini,
  Ward, Angiuoli, Crabtree, Jones, Durkin, DeBoy, Davidsen, Mora, Scarselli,
  Margarit, Peterson, Hauser, Sundaram, Nelson, Madupu, Brinkac, Dodson,
  Rosovitz, Sullivan, Daugherty, Haft, Selengut, Gwinn, Zhou, Zafar, Khouri,
  Radune, Dimitrov, Watkins, O'Connor, Smith, Utterback, White, Rubens, Grandi,
  Madoff, Kasper, Telford, Wessels, Rappuoli, and Fraser]{Tettelin2005}
Tettelin H, Masignani V, Cieslewicz MJ, Donati C, Medini D, Ward NL,
\emph{et al.} (2005).
\newblock Genome analysis of multiple pathogenic isolates of
  \emph{Streptococcus agalactiae}: Implications for the microbial
  ``pan-genome''.
\newblock \emph{Proc Natl Acad Sci USA} \textbf{102}: 13950--13955.

\bibitem[Torsvik \emph{et~al.}(1990)Torsvik, Salte, Sorheim, and Goksoyr]{Torsvik1990}
Torsvik V, Salte K, Sorheim R, Goksoyr J. (1990).
\newblock Comparison of phenotypic diversity and {DNA} heterogeneity in a
  population of soil bacteria.
\newblock \emph{Appl Environ Microbiol} \textbf{56}: 776--781.

\bibitem[Whitman \emph{et~al.}(1998)Whitman, Coleman, and Wiebe]{Whitman1998}
Whitman WB, Coleman DC, Wiebe WJ (1998).
\newblock Prokaryotes: The unseen majority.
\newblock \emph{Proc Natl Acad Sci USA} \textbf{95}: 6578--6583.

\bibitem[Wilson(1999)]{Wilson1999}
Wilson EO. (1999).
\newblock \emph{The Diversity of Life}.
\newblock W.W.~Norton \& Company.

\setcounter{firstbib}{\value{enumiv}}
\end{thebibliography}

\begin{thebibliography}{8}
\setcounter{enumiv}{\value{firstbib}}

\bibitem[Good(1953)]{Good1953supp}
Good IJ. (1953).  The population frequencies of species and the estimation of population parameters.  \emph{Biometrika} \textbf{40}: 237--264.

\bibitem[Huber \emph{et~al.}(2007)]{Huber2007supp}
Huber JA, Welch DBM, Morrison HG, Huse SM, Neal PR, Butterfield DA \emph{et al.} (2007).  Microbial population structures in the deep marine biosphere.  \emph{Science} \textbf{318}: 97--100.

\bibitem[Mao(2007)]{Mao2007supp}
Mao CX. (2007).  Estimating species accumulation curves and diversity indices.  \emph{Statist Sinica} \textbf{17}: 761--774.

\bibitem[N\'adas(1985)]{Nadas1985supp}
N\'adas A. (1985).  On {T}uring's formula for word probabilities.  \emph{IEEE Trans Acoust Speech Signal Processing} \textbf{33}: 1414--1416.

\bibitem[Quince \emph{et~al.}(2008)]{Quince2008supp}
Quince C, Curtis TP, Sloan WT. (2008).  The rational exploration of microbial diversity.  \emph{ISME J} \textbf{2}: 997--1006.

\bibitem[Roesch \emph{et~al.}(2007)]{Roesch2007supp}
Roesch LFW, Fulthorpe RR, Riva A, Casella G, Hadwin AKM, Kent AD \emph{et al.} (2007).  Pyrosequencing enumerates and contrasts soil microbial diversity.  \emph{ISME J} \textbf{1}: 283--290.

\bibitem[Rusch \emph{et~al.}(2007)]{Rusch2007supp}
Rusch DB, Halpern AL, Sutton G, Heidelberg KB, Williamson S, Yooseph S \emph{et al.} (2007).  The \emph{Sorcerer II} {G}lobal {O}cean {S}ampling expedition: {N}orthwest {A}tlantic through {E}astern {T}ropical {P}acific.  \emph{PLoS Biol} \textbf{5}: e77.

\bibitem[Tsallis(1988)]{Tsallis1988supp}
Tsallis C. (1988).  Possible generalization of {B}oltzmann-{G}ibbs statistics.  \emph{J Stat Phys} \textbf{52}: 479--487.
\end{thebibliography}
\end{document}